%% file: AudioInterference_arxiv.tex
\documentclass{article} 
\usepackage{iclr2027_conference,times}

\input{math_commands.tex}

\usepackage{booktabs}
\usepackage{array}
\usepackage[table]{xcolor}
\usepackage{hyperref}
\usepackage{url}
\usepackage{algorithm}
\usepackage{algpseudocode}
\usepackage{wrapfig}
\usepackage{multirow}
\usepackage{tabularx}
\usepackage{graphicx}

\definecolor{TableHeader}{HTML}{DCE6F1}
\definecolor{TableGroup}{HTML}{EEF2F6}
\definecolor{TableAlt}{HTML}{F7F9FB}
\definecolor{TablePositive}{HTML}{E8F4EA}
\definecolor{TableBoundary}{HTML}{FBEAEA}

\title{Selective Listening: Mechanism-Guided Control of Audio Influence in Large Audio-Language Models}

\author{
Yulin Sun\textsuperscript{1,2},
Kele Xu\textsuperscript{1,2}\thanks{Corresponding author.},
Yong Dou\textsuperscript{1,2}\\
\textsuperscript{1}College of Computer Science and Technology, National University of Defense Technology\\
\textsuperscript{2}State Key Laboratory of Complex \& Critical Software Environment
}

\iclrfinalcopy 
\begin{document}

\maketitle
\fancyhead{}
\renewcommand{\headrulewidth}{0pt}

\begin{abstract}
Large audio-language models (LALMs) exploit multimodal evidence, yet task-irrelevant audio can alter text-reasoning decisions when listening is unnecessary. Aggregate Accuracy can hide this paired drift because audio-induced repairs and damages may cancel. Paired drift analysis and targeted interventions identify architecture-specific, intervention-sensitive late audio pathways as actionable control points. We introduce ICAP-Gate, which applies mechanism-guided, task-conditioned control to each model's pathway. Across four LALMs, two reasoning benchmarks, and environmental-sound and natural-speech interference, ICAP-Gate has lower point estimates for Influence Rate and Answer Flip than ungated inference in all 16 full-split model--condition evaluations. Fixed suppression degrades automatic speech recognition (ASR) across all four models, whereas ICAP-Gate matches ungated ASR performance by preserving the pathway for explicit audio-demand instructions. ICAP-Gate has lower paired-drift point estimates than mitigation prompting in all four evaluated settings and provides competitive stabilization relative to eight-sample Self-Consistency while using one generation per query; in controlled ARC measurements, Self-Consistency incurs $7.0$--$9.2\times$ ungated latency. These results establish \emph{selective modality influence control} as a design principle for robust multimodal reasoning.
\end{abstract}

\section{Introduction}
\label{sec:introduction}

\begin{figure}[t]
    \centering
    \includegraphics[width=0.90\linewidth]{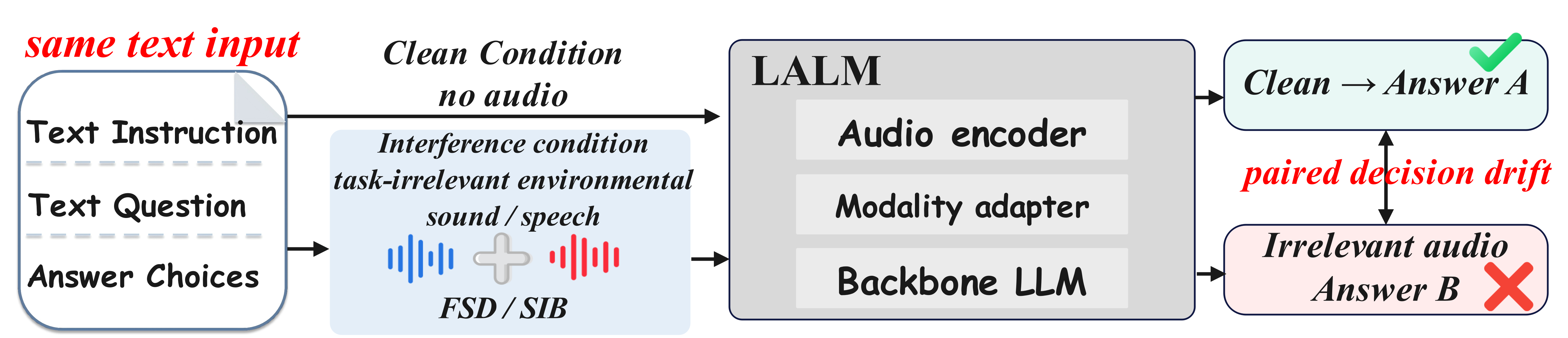}
    \caption{\textbf{Paired decision drift under task-irrelevant audio.} The same text input is evaluated with and without task-irrelevant environmental sound or speech; the downstream decision may change.}
    \label{fig:phenomenon-teaser}
\end{figure}

Large audio-language models (LALMs) combine listening with language reasoning, but modality availability does not imply task relevance. In text-reasoning settings, environmental sounds or speech may provide no evidence for the answer yet change the model's decision. Robust multimodal inference therefore requires control over when audio influences computation. Figure~\ref{fig:phenomenon-teaser} illustrates this paired decision drift.

Recent work shows that irrelevant audio can degrade text reasoning and increase output volatility in LALMs \citep{li2026silence}. Aggregate task scores can conceal that instability when repairs and damages cancel. We therefore measure \emph{paired decision drift}: whether the same text question reaches a different decision with task-irrelevant audio.

Behavioral drift does not reveal where irrelevant audio acquires consequential influence inside the model or identify a control point that can be modulated according to task demand. Because the audio pathway must remain available for explicit audio-demand instructions and the evaluated audio-relevant tasks, the mechanistic question is: \emph{where does irrelevant audio become actionable in reasoning, and can its influence be selectively controlled while preserving useful audio processing?}


We trace unwanted modality influence from behavior through mechanism to control. Paired clean/interference evaluation and targeted internal interventions identify intervention-sensitive late audio pathways through which irrelevant audio affects downstream decisions. Qwen2.5-Omni provides the highest-resolution causal localization, while actionable late-path control points recur in Phi-4-MM and Voxtral-Mini-3B. This pattern motivates \emph{selective modality influence control}, instantiated by ICAP-Gate: it modulates each model's pathway according to task demand with one generation per query, stabilizing reasoning across architectures and interference types while preserving performance on the evaluated ASR task. Figure~\ref{fig:influence-to-control} summarizes this mechanism-guided control.

Our contributions are threefold:
\begin{itemize}
    \item \textbf{Paired drift exposes the hidden failure mode.} Paired decision analysis reveals reasoning drift hidden by aggregate Accuracy, while targeted interventions causally localize an intervention-sensitive late audio pathway in Qwen2.5-Omni.
    
    \item \textbf{Late audio pathways provide actionable control points.} We formulate \emph{selective modality influence control} and instantiate it with ICAP-Gate, which conditions architecture-adapted pathway modulation on task demand so that audio influence is attenuated when unnecessary and preserved for explicit audio-demand instructions. Corresponding actionable sites recur across heterogeneous audio architectures.
    
    \item \textbf{Task-conditioned control stabilizes reasoning while preserving evaluated ASR.} Across four LALMs, two reasoning benchmarks, and both environmental and controlled natural-speech interference, ICAP-Gate has lower Influence Rate and Answer Flip point estimates than ungated inference in all 16 full-split evaluations while matching ungated WER on the evaluated ASR task.
\end{itemize}

\section{Related Work}
\label{sec:related-work}

\subsection{Large Audio-Language Models and Multimodal Reasoning}
\label{sec:rw-lalms}

Audio-language models extend language reasoning to audio understanding and spoken interaction. Early systems linked pretrained language models to audio and speech representations, while later systems broadened audio instruction tuning, long-context reasoning, and end-to-end speech interaction \citep{deshmukh2023pengi,gong2024listen,zhang2023speechgpt,rubenstein2023audiopalm,tang2024salmonn,ghosh2025audio,defossez2024moshi,huang2025step,li2025baichuan}.

Recent systems expose distinct audio--language interfaces, including Qwen2.5-Omni, Phi-4-Multimodal, and Voxtral \citep{xu2025qwen25omnitechnicalreport,abouelenin2025phi,liu2025voxtral,chu2023qwen,chu2024qwen2}. This architectural diversity sharpens the control question: how should an LALM regulate audio influence when listening is unnecessary?

\subsection{Robustness to Irrelevant or Distracting Modalities}
\label{sec:rw-distraction}

Irrelevant textual, retrieved, visual, and multimodal contexts can disrupt reasoning across language and multimodal models \citep{shi2023large,yoran2024making,yang2025llm,sharma2024losing,yang2026defying,cai2025diagnosing,liu2024insight,tian2026crosscheck}. Multimodal capability alone therefore does not ensure selective use of available inputs.

For audio-language models, \citet{li2026silence} showed that task-irrelevant silence, synthetic noise, and environmental sounds perturb text reasoning across multiple LALMs, and evaluated prompting and Self-Consistency as inference-time mitigations. We identify where irrelevant audio becomes consequential and control that pathway according to task demand.

\subsection{Mechanistic Intervention and Conditional Modality Control}
\label{sec:rw-mechanism-control}

Mechanistic interventions—including causal mediation and tracing, activation patching, circuit discovery, and representation patching—localize behaviorally consequential computation \citep{vig2020investigating,meng2022locating,conmy2023towards,zhang2024towards,ghandeharioun2024patchscopes}. Multimodal studies apply related interventions to vision--language representations and information flow \citep{palit2023towards,li2026causal,kim2026map}, establishing internal pathways as actionable objects for explanation and control.

Conditional computation regulates information flow through gating, feature-wise conditioning, cross-attention, and expert routing \citep{arevalo2017gated,perez2018film,alayrac2022flamingo,shazeer2017outrageously,fedus2022switch}; multimodal routing extends this principle across modalities \citep{wu2024omni,lin2026moe}. We connect intervention-based pathway localization with task-conditioned control: ICAP-Gate modulates the architecture-specific pathway through which irrelevant audio affects reasoning, realizing \emph{selective modality influence control}.

\begin{figure}[t]
    \centering
    \includegraphics[width=\textwidth]{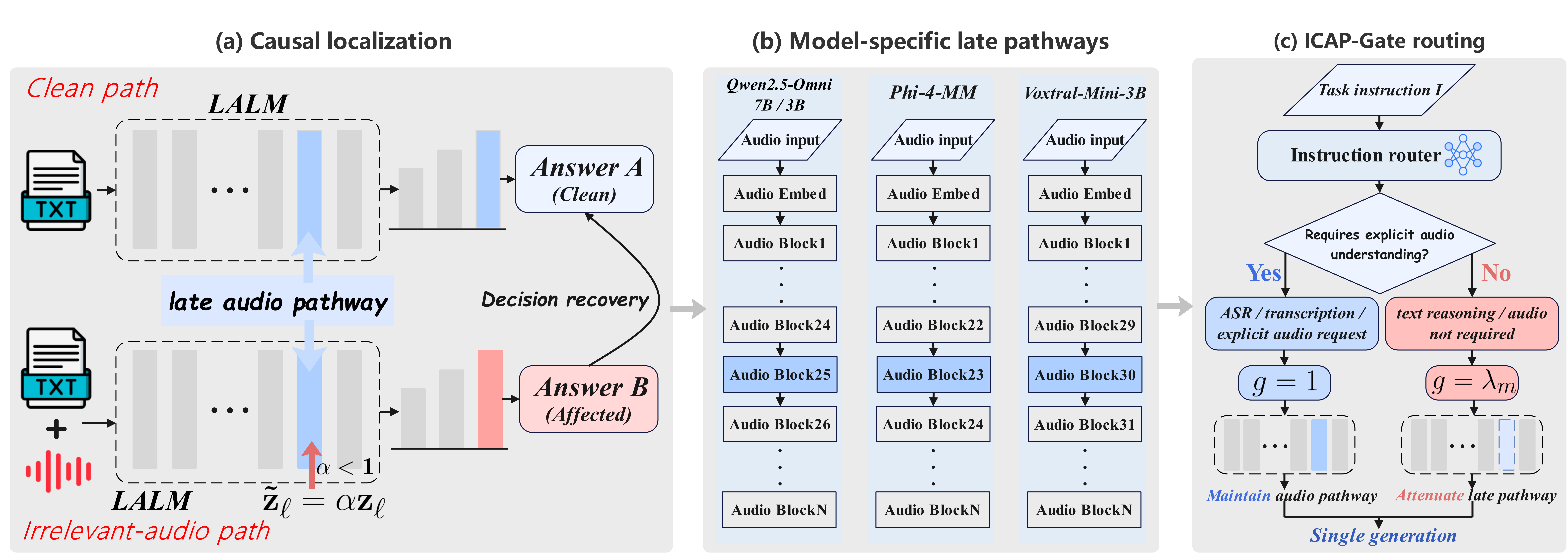}
    \caption{\textbf{Mechanism-guided selective control of irrelevant-audio influence in LALMs.} \textbf{(a)} Pathway scaling identifies a late audio pathway through which irrelevant audio changes Qwen2.5-Omni decisions. \textbf{(b)} The actionable control site is architecture-adapted across Qwen2.5-Omni, Phi-4-MM, and Voxtral-Mini-3B. \textbf{(c)} Instruction-conditioned routing keeps the pathway open for explicit audio-demand requests ($g=1$) and attenuates it otherwise ($g=\lambda_m$), followed by one generation.}
    \label{fig:influence-to-control}
\end{figure}

\section{Methodology}
\label{sec:methodology}

\subsection{Problem Formulation and Reasoning Drift}
\label{sec:problem-formulation}

We study \emph{unwanted modality influence} in large audio-language models: audio should affect a prediction when it provides task-relevant evidence, but its presence alone should not alter a decision when the task can be solved from text. Let $\mathcal{D}=\{(x_i,y_i)\}_{i=1}^{N}$ denote a text-reasoning dataset, where $x_i$ is the reasoning input and $y_i$ is the gold answer. Each example is evaluated under three aligned conditions: clean text-only inference, ungated inference with task-irrelevant audio $a_i$, and pathway-controlled inference. For the interference conditions considered, $y_i$ is determined by $x_i$, while $a_i$ provides no answer-bearing information. We instantiate $a_i$ with structured environmental audio (FSD) and natural speech (SIB), covering non-linguistic and linguistically structured interference.

For a model $f_{\theta}$ and answer extractor $E$, the corresponding predictions are
\begin{equation}
\hat{y}^{\,c}_i = E\!\left(f_{\theta}(x_i)\right), \qquad
\hat{y}^{\,u}_i = E\!\left(f_{\theta}(x_i,a_i)\right), \qquad
\hat{y}^{\,g}_i = E\!\left(f_{\theta,G}(x_i,a_i)\right),
\label{eq:prediction-conditions}
\end{equation}
where $c$, $u$, and $g$ denote clean, ungated, and pathway-controlled inference, respectively, and $G$ denotes the pathway-control intervention. Clean inference provides the paired reference for isolating audio-induced change, while correctness is tracked independently against the gold answer. We operationalize \emph{reasoning drift} as a sample-level change in either correctness state or extracted answer identity relative to this clean reference.

For \(k\in\{c,u,g\}\), define correctness and accuracy as
\begin{equation}
s_i^{k}=\mathbf{1}\!\left[\hat{y}^{\,k}_i=y_i\right],
\qquad
\mathrm{Acc}(k)=\frac{1}{N}\sum_{i=1}^{N}s_i^{k}.
\label{eq:accuracy}
\end{equation}
We measure paired drift using Influence Rate (IR) for correctness-state changes and Answer Flip for any answer change, including switches between incorrect answers:
\begin{equation}
\mathrm{IR}(k)=
\frac{1}{N}\sum_{i=1}^{N}
\mathbf{1}\!\left[s_i^{k}\neq s_i^{c}\right],
\qquad
\mathrm{Flip}(k)=
\frac{1}{N}\sum_{i=1}^{N}
\mathbf{1}\!\left[\hat{y}^{\,k}_i\neq\hat{y}^{\,c}_i\right].
\label{eq:drift-metrics}
\end{equation}

The distinction from aggregate accuracy is exact. Let $n_{\mathrm{cw}}^{(k)}$ and $n_{\mathrm{wc}}^{(k)}$ denote clean-correct$\rightarrow$wrong and clean-wrong$\rightarrow$correct transitions under condition $k$. Then
\begin{equation}
\mathrm{IR}(k)=
\frac{n_{\mathrm{cw}}^{(k)}+n_{\mathrm{wc}}^{(k)}}{N},
\qquad
\mathrm{Acc}(k)-\mathrm{Acc}(c)=
\frac{n_{\mathrm{wc}}^{(k)}-n_{\mathrm{cw}}^{(k)}}{N}.
\label{eq:accuracy-drift-decomposition}
\end{equation}
Accuracy can therefore remain nearly unchanged even when many individual predictions move, because opposing transitions cancel in aggregate. We consequently report accuracy together with paired drift rather than treating aggregate score as a complete measure of robustness.

Finally, we resolve the \emph{direction} of a pathway intervention with two paired transition measures:
\begin{alignat}{2}
&\mathrm{NetC2W} \quad &&=
\sum_{i=1}^{N}\mathbf{1}
[s_i^{c}=1,s_i^{u}=0,s_i^{g}=1]
-
\sum_{i=1}^{N}\mathbf{1}
[s_i^{c}=1,s_i^{u}=1,s_i^{g}=0],
\label{eq:net-c2w}\\
&\mathrm{NetAnswer} \quad &&=
\sum_{i=1}^{N}\mathbf{1}
[\hat{y}^{\,c}_i\neq\hat{y}^{\,u}_i,\,
 \hat{y}^{\,c}_i=\hat{y}^{\,g}_i]
-
\sum_{i=1}^{N}\mathbf{1}
[\hat{y}^{\,c}_i=\hat{y}^{\,u}_i,\,
 \hat{y}^{\,c}_i\neq\hat{y}^{\,g}_i].
\label{eq:net-answer}
\end{alignat}
$\mathrm{NetC2W}$ measures net recovery of clean-correct cases, while $\mathrm{NetAnswer}$ measures net restoration of the clean decision. Together with IR and Answer Flip, these paired quantities expose whether pathway control reverses the specific prediction changes introduced by irrelevant audio.

\subsection{Late Audio Pathways Provide Actionable Control Points}
\label{sec:pathway-localization}

The paired drift measures establish \emph{when} irrelevant audio changes a reasoning decision; we next ask \emph{where} that influence becomes actionable. Paired traces and teacher-forced answer analysis identify where executions separate, while targeted interventions test whether the pathway controls the resulting transitions. Figure~\ref{fig:influence-to-control}(a) summarizes this localization, which is defined by its effect on paired reasoning transitions rather than activation magnitude alone.

For an audio-pathway module at depth $\ell$ with output $\mathbf{z}_{\ell}$, we intervene using
\begin{equation}
    \widetilde{\mathbf{z}}_{\ell}
    =
    \alpha \mathbf{z}_{\ell},
    \qquad 0 \leq \alpha \leq 1,
    \label{eq:pathway-intervention}
\end{equation}
where $\alpha=0$ corresponds to ablation and $0<\alpha<1$ to soft attenuation. We quantify intervention sensitivity with Influence Rate, Answer Flip, Net C2W, Net Answer, and the gold-answer margin, identifying pathway sites whose modulation changes downstream reasoning.

\begin{wrapfigure}[20]{r}{0.40\textwidth}
\vspace{-0.35\baselineskip}
\centering
\includegraphics[width=\linewidth]{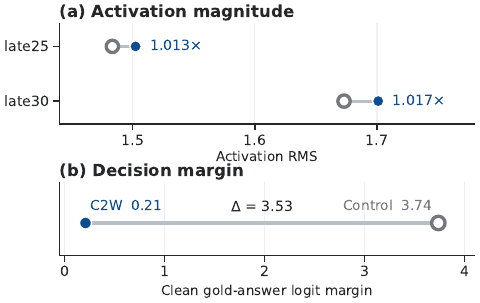}
\setlength{\abovecaptionskip}{2pt}
\caption{\textbf{Activation magnitude and decision margin (Qwen2.5-Omni, FSD).} \textbf{(a)} Activation RMS at two late audio sites for clean-correct-to-wrong (C2W) and clean-stable controls; labels show C2W/control ratios. \textbf{(b)} Shared clean gold-answer logit margins for groups. Filled blue/hollow gray markers denote C2W/control aggregates.}
\label{fig:activation-margin-contrast}
\vspace{-0.25\baselineskip}
\end{wrapfigure}

In Qwen2.5-Omni, complementary diagnostics and interventions identify a late audio pathway as an actionable control point. Intervening at this pathway reverses audio-induced answer changes and restores clean decisions, showing that it controls paired reasoning transitions. Figure~\ref{fig:activation-margin-contrast} shows that clean-correct-to-wrong and clean-stable cases have similar audio-activation RMS at representative late sites, while their clean gold-answer margins differ sharply. Together with the targeted interventions, this contrast points to decision-relevant influence through the late audio pathway rather than a simple increase in activation magnitude.

The same intervention logic recurs across heterogeneous LALM architectures. Phi-4-MM and Voxtral-Mini-3B expose model-specific late audio pathways with the same actionable control role, as summarized in Figure~\ref{fig:influence-to-control}(b). The module and attenuation scale are architecture-specific, while the higher-level pattern is consistent: unwanted audio influence remains controllable through late audio processing. Detailed Qwen localization evidence is provided in Appendix~\ref{app:qwen-mechanistic-localization}.

This localization turns mitigation into internal control of audio \emph{influence}, rather than removal of the audio input.

\subsection{ICAP-Gate Conditions Pathway Influence by Task Demand}
\label{sec:icap-gate}

The localization shows that the audio pathway can carry unwanted influence when audio is irrelevant yet remains important for explicit audio-demand instructions and the evaluated ASR task. We therefore use \emph{selective modality influence control}: Figure~\ref{fig:influence-to-control}(c) instantiates this principle as \textbf{ICAP-Gate} (Instruction-Conditioned Audio Pathway Gating), which modulates the intervention-sensitive pathway identified for each architecture.

For model $m$, let $\ell_m$ denote the selected audio pathway and $\lambda_m\in[0,1)$ its calibrated attenuation scale. Given task instruction $I$, we define an explicit audio-demand indicator $r(I)\in\{0,1\}$ and apply
\begin{equation}
    g_m(I)=
    \begin{cases}
        1, & r(I)=1,\\
        \lambda_m, & r(I)=0,
    \end{cases}
    \qquad
    \widetilde{\mathbf{z}}_{\ell_m}
    =
    g_m(I)\mathbf{z}_{\ell_m}.
    \label{eq:icap-gate}
\end{equation}
When the instruction explicitly asks for audio understanding, ICAP-Gate leaves the pathway unchanged; otherwise, it applies the model-specific attenuation. The rule is therefore architecture-adapted rather than tied to a particular layer index.

ICAP-Gate routes on the \emph{task instruction}, whose explicit audio-demand cue determines $r(I)$, rather than on incidental words in the question body. The cue inventory and routing audit are provided in Appendix~\ref{app:routing-audit}; instruction-scoped routing keeps the decision transparent and auditable.

We obtain $(\ell_m,\lambda_m)$ offline from the intervention analysis and report the selected pathway and full-split confirmation scope in Table~\ref{tab:pathway-summary}. Online inference requires one instruction routing decision and one temporary pathway modulation, preserving single-generation inference without a clean-reference pass or auxiliary relevance model. The complete procedure is given in Appendix~\ref{app:icap-inference}, Algorithm~\ref{alg:icap-gate}.

\section{Experiments}
\label{sec:experiments}

\subsection{Experimental Setup}
\label{sec:experimental-setup}

\paragraph{Models and reasoning tasks.}
We evaluate four LALMs---Qwen2.5-Omni-7B, Qwen2.5-Omni-3B, Phi-4-MM, and Voxtral-Mini-3B---on the full ARC-Challenge and MMLU test sets. Each benchmark is paired with FSD and SIB, yielding four full-split settings per model. Table~\ref{tab:pathway-summary} reports the architecture-specific intervention configurations; the control rule is shared across models.

\paragraph{Irrelevant-audio conditions.}
FSD pairs each reasoning example with a real-world environmental clip drawn from FSD50K \citep{fonseca2022fsd50k}. SIB complements this non-linguistic interference with independently paired natural speech from LibriSpeech \texttt{dev-clean} \citep{panayotov2015librispeech}, while the gold answer remains determined by the text. This tests whether unwanted modality influence persists when the irrelevant stream carries coherent linguistic content; construction and transcript validity are detailed in Section~\ref{sec:sib-validity} and Appendix~\ref{app:sib-validity}.

\paragraph{Inference and evaluation protocol.}
For every model and example, clean, ungated-audio, and ICAP-Gate predictions use the same text input, paired interference audio, and model-specific decoding configuration. Detailed response handling and source-level alignment are provided in Appendix~\ref{app:reproducibility}.

\paragraph{Statistics and audio-relevant evaluation.}
We report Accuracy, Influence Rate, Answer Flip, NetC2W, and NetAnswer as defined in Section~\ref{sec:problem-formulation}; statistical checks and implementation details are provided in Appendices~\ref{app:reproducibility} and~\ref{app:statistical-analysis}. Selective preservation is evaluated with ungated, fixed-attenuation, and ICAP-Gate inference on 400 LibriSpeech \texttt{dev-clean} ASR utterances per model using word error rate (WER).

\subsection{ICAP-Gate Lowers Paired-Drift Point Estimates Across Four LALMs}
\label{sec:main-results}

ICAP-Gate has lower point estimates of both paired drift measures than ungated inference in all 16 full-split model--condition evaluations. Table~\ref{tab:main-results} reports the complete matrix, while Figure~\ref{fig:selective-control-summary}(b) summarizes the corresponding IR reductions as a cross-model heatmap. The exact intervention pathway is architecture-specific, yet selective control of that pathway repeatedly makes text reasoning less sensitive to an audio stream that is unnecessary for solving the task.

\begin{table}[t]
\centering
\caption{\textbf{Main results across four LALMs.} Accuracy is reported as clean/ungated/ICAP (C/U/I); IR and Flip show ungated$\rightarrow$ICAP. $\Delta$IR and $\Delta$Flip denote absolute reductions in percentage points. Net C2W and Net Ans report directional recovery toward the clean decision. Asterisks mark paired 95\% CIs that exclude zero.}
\label{tab:main-results}
\small
\setlength{\tabcolsep}{4.0pt}
\renewcommand{\arraystretch}{1.00}
\newcommand{\mainsig}{\rlap{$^{\ast}$}}

\begin{tabular}{lcccccrr}
\toprule
\rowcolor{black!10}
\textbf{Setting} &
\textbf{Acc. C/U/I} &
\textbf{IR U$\rightarrow$I} &
\textbf{$\Delta$IR$\uparrow$} &
\textbf{Flip U$\rightarrow$I} &
\textbf{$\Delta$Flip$\uparrow$} &
\textbf{Net C2W} &
\textbf{Net Ans} \\
\midrule

\rowcolor{black!7}
\multicolumn{8}{l}{\textbf{Qwen2.5-Omni-7B}} \\
ARC-FSD
& 84.56/83.53/\textbf{84.90}
& 8.19$\rightarrow$\textbf{6.14}
& \cellcolor{TablePositive}\textbf{2.05}\mainsig
& 9.04$\rightarrow$\textbf{7.00}
& \cellcolor{TablePositive}\textbf{2.04}\mainsig
& \textbf{+20}
& \textbf{+24} \\
ARC-SIB
& 84.56/82.85/\textbf{85.07}
& 9.39$\rightarrow$\textbf{7.17}
& \cellcolor{TablePositive}\textbf{2.22}\mainsig
& 11.09$\rightarrow$\textbf{8.28}
& \cellcolor{TablePositive}\textbf{2.81}\mainsig
& \textbf{+26}
& \textbf{+33} \\
MMLU-FSD
& 67.42/67.58/\textbf{67.55}
& 11.80$\rightarrow$\textbf{9.16}
& \cellcolor{TablePositive}\textbf{2.64}\mainsig
& 16.06$\rightarrow$\textbf{12.60}
& \cellcolor{TablePositive}\textbf{3.46}\mainsig
& \textbf{+183}
& \textbf{+486} \\
MMLU-SIB
& 67.42/66.00/\textbf{67.68}
& 13.65$\rightarrow$\textbf{11.19}
& \cellcolor{TablePositive}\textbf{2.46}\mainsig
& 18.89$\rightarrow$\textbf{15.58}
& \cellcolor{TablePositive}\textbf{3.31}\mainsig
& \textbf{+291}
& \textbf{+465} \\

\addlinespace[2pt]
\rowcolor{black!7}
\multicolumn{8}{l}{\textbf{Qwen2.5-Omni-3B}} \\
ARC-FSD
& 77.39/77.56/\textbf{77.13}
& 11.77$\rightarrow$\textbf{7.76}
& \cellcolor{TablePositive}\textbf{4.01}\mainsig
& 13.82$\rightarrow$\textbf{9.73}
& \cellcolor{TablePositive}\textbf{4.09}\mainsig
& \textbf{+21}
& \textbf{+48} \\
ARC-SIB
& 77.39/76.54/\textbf{76.62}
& 13.82$\rightarrow$\textbf{10.84}
& \cellcolor{TablePositive}\textbf{2.98}\mainsig
& 16.13$\rightarrow$\textbf{12.63}
& \cellcolor{TablePositive}\textbf{3.50}\mainsig
& \textbf{+18}
& \textbf{+41} \\
MMLU-FSD
& 60.36/59.76/\textbf{60.25}
& 15.24$\rightarrow$\textbf{13.41}
& \cellcolor{TablePositive}\textbf{1.83}\mainsig
& 21.61$\rightarrow$\textbf{19.40}
& \cellcolor{TablePositive}\textbf{2.21}\mainsig
& \textbf{+163}
& \textbf{+310} \\
MMLU-SIB
& 60.36/58.97/\textbf{59.78}
& 17.79$\rightarrow$\textbf{14.29}
& \cellcolor{TablePositive}\textbf{3.50}\mainsig
& 25.17$\rightarrow$\textbf{20.65}
& \cellcolor{TablePositive}\textbf{4.52}\mainsig
& \textbf{+303}
& \textbf{+635} \\

\addlinespace[2pt]
\rowcolor{black!7}
\multicolumn{8}{l}{\textbf{Phi-4-MM (5.6B)}} \\
ARC-FSD
& 81.66/75.51/\textbf{75.77}
& 22.53$\rightarrow$\textbf{21.93}
& \cellcolor{TablePositive}\textbf{0.60}
& 25.00$\rightarrow$\textbf{24.66}
& \cellcolor{TablePositive}\textbf{0.34}
& \textbf{+5}
& \textbf{+4} \\
ARC-SIB
& 81.66/73.21/\textbf{75.51}
& 23.46$\rightarrow$\textbf{21.33}
& \cellcolor{TablePositive}\textbf{2.13}
& 26.54$\rightarrow$\textbf{24.66}
& \cellcolor{TablePositive}\textbf{1.88}
& \textbf{+26}
& \textbf{+22} \\
MMLU-FSD
& 63.99/57.28/\textbf{58.98}
& 29.38$\rightarrow$\textbf{28.04}
& \cellcolor{TablePositive}\textbf{1.34}\mainsig
& 39.56$\rightarrow$\textbf{38.04}
& \cellcolor{TablePositive}\textbf{1.52}\mainsig
& \textbf{+213}
& \textbf{+214} \\
MMLU-SIB
& 63.99/57.25/\textbf{57.51}
& 30.00$\rightarrow$\textbf{28.96}
& \cellcolor{TablePositive}\textbf{1.04}\mainsig
& 40.51$\rightarrow$\textbf{38.73}
& \cellcolor{TablePositive}\textbf{1.78}\mainsig
& \textbf{+92}
& \textbf{+249} \\

\addlinespace[2pt]
\rowcolor{black!7}
\multicolumn{8}{l}{\textbf{Voxtral-Mini-3B}} \\
ARC-FSD
& 75.77/73.38/\textbf{71.76}
& 21.33$\rightarrow$\textbf{20.90}
& \cellcolor{TablePositive}\textbf{0.43}
& 26.02$\rightarrow$\textbf{25.85}
& \cellcolor{TablePositive}\textbf{0.17}
& \textbf{-7}
& \textbf{+2} \\
ARC-SIB
& 75.77/69.88/\textbf{73.38}
& 25.17$\rightarrow$\textbf{23.04}
& \cellcolor{TablePositive}\textbf{2.13}
& 30.20$\rightarrow$\textbf{27.22}
& \cellcolor{TablePositive}\textbf{2.98}\mainsig
& \textbf{+33}
& \textbf{+35} \\
MMLU-FSD
& 57.79/58.00/\textbf{58.21}
& 27.44$\rightarrow$\textbf{26.26}
& \cellcolor{TablePositive}\textbf{1.18}\mainsig
& 37.53$\rightarrow$\textbf{36.00}
& \cellcolor{TablePositive}\textbf{1.53}\mainsig
& \textbf{+98}
& \textbf{+215} \\
MMLU-SIB
& 57.79/56.62/\textbf{58.76}
& 28.41$\rightarrow$\textbf{25.88}
& \cellcolor{TablePositive}\textbf{2.53}\mainsig
& 39.52$\rightarrow$\textbf{36.28}
& \cellcolor{TablePositive}\textbf{3.24}\mainsig
& \textbf{+328}
& \textbf{+454} \\
\bottomrule
\end{tabular}
\end{table}

\begin{table}[t]
\centering
\footnotesize
\caption{\textbf{Measured test-time cost.} Statistics aggregate 200 measured queries and three sequential repeats after 20 warm-up queries on one NVIDIA A800 80GB GPU (bfloat16; model loading excluded). VRAM is peak allocated memory. Green marks ICAP-Gate; red marks Self-Consistency.}
\label{tab:inference-cost}
\setlength{\tabcolsep}{3.5pt}
\renewcommand{\arraystretch}{1.12}
\begin{tabularx}{\textwidth}{c>{\raggedright\arraybackslash}Xcccccc}
\toprule
\rowcolor{TableHeader}
\textbf{Dataset} & \textbf{Method} & \textbf{Gen./q.} & \textbf{Mean (s)} & \textbf{Median (s)} & \textbf{Rel.} & \textbf{VRAM (GiB)} & \textbf{Tok./q.} \\
\midrule
\multirow[c]{4}{*}{\textbf{ARC-FSD}} & Ungated & 1 & 6.8881 & 6.4671 & 1.000 & 17.03 & 85.8 \\
& Prompting & 1 & 6.6079 & 5.8928 & 0.959 & 17.04 & 81.2 \\
& \cellcolor{TablePositive}\textbf{ICAP-Gate} & \cellcolor{TablePositive}\textbf{1} & \cellcolor{TablePositive}\textbf{7.2840} & \cellcolor{TablePositive}\textbf{6.6174} & \cellcolor{TablePositive}\textbf{1.057} & \cellcolor{TablePositive}\textbf{17.03} & \cellcolor{TablePositive}\textbf{94.3} \\
& \cellcolor{TableBoundary}\textbf{Self-Consistency} & \cellcolor{TableBoundary}\textbf{8} & \cellcolor{TableBoundary}63.4709 & \cellcolor{TableBoundary}57.7551 & \cellcolor{TableBoundary}\textbf{9.215} & \cellcolor{TableBoundary}17.03 & \cellcolor{TableBoundary}\textbf{819.4} \\
\midrule
\multirow[c]{4}{*}{\textbf{ARC-SIB}} & Ungated & 1 & 8.8734 & 8.0748 & 1.000 & 16.85 & 88.8 \\
& Prompting & 1 & 7.9270 & 7.1944 & 0.893 & 16.85 & 81.6 \\
& \cellcolor{TablePositive}\textbf{ICAP-Gate} & \cellcolor{TablePositive}\textbf{1} & \cellcolor{TablePositive}\textbf{7.9317} & \cellcolor{TablePositive}\textbf{7.4159} & \cellcolor{TablePositive}\textbf{0.894} & \cellcolor{TablePositive}\textbf{16.85} & \cellcolor{TablePositive}\textbf{90.2} \\
& \cellcolor{TableBoundary}\textbf{Self-Consistency} & \cellcolor{TableBoundary}\textbf{8} & \cellcolor{TableBoundary}62.3897 & \cellcolor{TableBoundary}53.6749 & \cellcolor{TableBoundary}\textbf{7.031} & \cellcolor{TableBoundary}16.85 & \cellcolor{TableBoundary}\textbf{817.8} \\
\bottomrule
\end{tabularx}
\end{table}

Paired metrics reveal the effect beyond aggregate Accuracy. On Qwen2.5-Omni-7B MMLU-FSD, Accuracy remains essentially unchanged (67.58\%$\rightarrow$67.55\%), while IR drops from 11.80\% to 9.16\% and Answer Flip from 16.06\% to 12.60\%. Across the four LALMs, the point estimates for IR and Answer Flip are lower than under ungated inference in all 16 settings. The natural-speech conditions show the same point-estimate direction across architectures, extending the stabilization result beyond environmental sounds.

\subsection{ICAP-Gate Provides Competitive Stability Within the Single-Generation Regime}
\label{sec:mitigation-efficiency}

\begin{wrapfigure}[19]{r}{0.56\textwidth}
\vspace{-0.45\baselineskip}
\centering
\includegraphics[width=\linewidth,trim=0 6pt 0 0,clip]{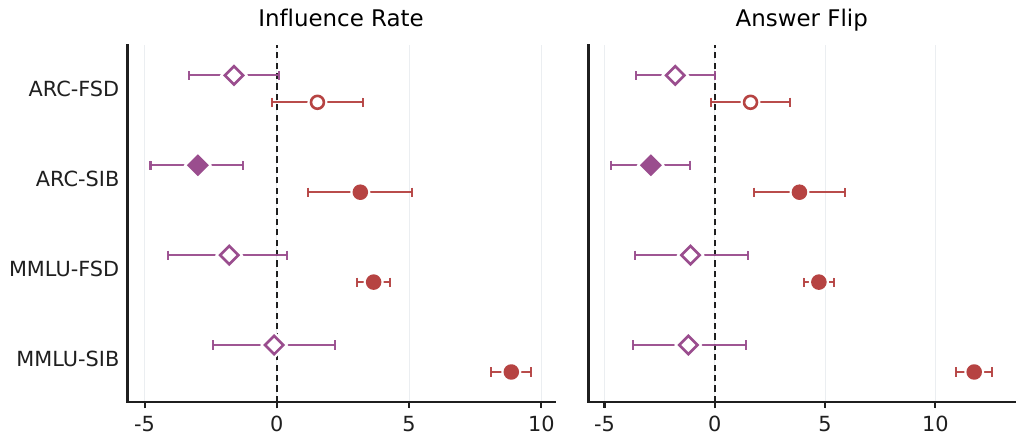}
\setlength{\abovecaptionskip}{2pt}
\caption{\textbf{Matched drift relative to ICAP-Gate.} Circles and diamonds denote Prompting and Self-Consistency, respectively. Values are comparator minus ICAP-Gate in percentage points: positive values indicate lower drift under ICAP-Gate, whereas negative values indicate lower drift under the comparator. Error bars are paired 95\% confidence intervals, with filled markers denoting intervals excluding zero.}
\label{fig:matched-drift}
\vspace{-0.25\baselineskip}
\end{wrapfigure}

We compare mechanism-guided pathway control with two inference-time strategies from prior irrelevant-audio studies~\citep{li2026silence}: input-level prompting and eight-sample Self-Consistency decoding. Prompting operates at the input level, Self-Consistency aggregates outputs, and ICAP-Gate scales the late audio pathway identified by the mechanistic analysis. Prompting and ICAP-Gate use one generation per query, whereas Self-Consistency aggregates eight responses.

All pairwise comparisons use the same Qwen2.5-Omni-7B checkpoint and aligned source examples within each comparison. Because mitigation strategies can change clean predictions, Influence Rate and Answer Flip are computed relative to each method's own clean reference. Matched-split sizes and protocol details are provided in Appendix~\ref{app:matched-mitigation}.

\begin{wraptable}{r}{0.60\textwidth}
\vspace{-0.55\baselineskip}
\centering
\footnotesize
\caption{\textbf{Matched drift relative to ICAP-Gate} (percentage points). Values are comparator minus ICAP-Gate: positive values favor ICAP-Gate, whereas negative values favor the comparator. Bold entries denote paired differences whose 95\% confidence intervals exclude zero (20,000 paired bootstrap resamples; seed 0), irrespective of which method is favored.}
\label{tab:mitigation-comparison}
\setlength{\tabcolsep}{2.0pt}
\renewcommand{\arraystretch}{1.08}
\begin{tabular*}{\linewidth}{c@{\extracolsep{\fill}}rrrr}
\toprule
\multirow[c]{2}{*}{\textbf{Setting}} & \multicolumn{2}{c}{\textbf{\shortstack[c]{Prompt--ICAP\\(1 vs.\ 1 gen.)}}} & \multicolumn{2}{c}{\textbf{\shortstack[c]{SCS--ICAP\\(8 vs.\ 1 gen.)}}} \\
\cmidrule[\heavyrulewidth]{2-5}
& \textbf{$\Delta$IR} & \textbf{$\Delta$Flip} & \textbf{$\Delta$IR} & \textbf{$\Delta$Flip} \\
\midrule
ARC-FSD & $+1.54$ & $+1.62$ & $-1.62$ & $-1.79$ \\
ARC-SIB & $\mathbf{+3.16}$ & $\mathbf{+3.84}$ & $\mathbf{-2.99}$ & $\mathbf{-2.90}$ \\
MMLU-FSD & $\mathbf{+3.66}$ & $\mathbf{+4.72}$ & $-1.80$ & $-1.10$ \\
MMLU-SIB & $\mathbf{+8.87}$ & $\mathbf{+11.76}$ & $-0.10$ & $-1.20$ \\
\bottomrule
\end{tabular*}
\vspace{-0.25\baselineskip}
\end{wraptable}

The mitigation prompt had higher IR and Answer Flip point estimates than ICAP-Gate in all four evaluated settings; on MMLU-SIB, the differences were $+8.87$ and $+11.76$ percentage points, with paired intervals excluding zero. Input-level guidance therefore does not directly control the identified pathway or reproduce the paired stability obtained through pathway control (Table~\ref{tab:mitigation-comparison}; Figure~\ref{fig:matched-drift}).

Self-Consistency had lower IR and Answer Flip point estimates than ICAP-Gate in all four matched comparisons, with its clearest paired advantage on ARC-SIB (Table~\ref{tab:mitigation-comparison}; Appendix~\ref{app:matched-mitigation}). It used eight generations per query. ICAP-Gate therefore provides competitive decision stability within the single-generation regime.

We further quantified the test-time cost of each inference path under a fixed single-GPU protocol. ICAP-Gate uses one generation per query and requires $0.894$--$1.057\times$ the Ungated latency across ARC-FSD and ARC-SIB. Self-Consistency uses eight generations and requires $7.031$--$9.215\times$ the Ungated latency. Detailed timing, token, and VRAM statistics are reported in Table~\ref{tab:inference-cost}.

Together, the comparisons place ICAP-Gate's advantage in mechanism-guided stability at single-generation cost: it directly controls the identified pathway, whereas prompting acts at the input and Self-Consistency uses repeated decoding. This reduces unwanted modality influence while preserving audio processing for explicit audio-demand instructions and the evaluated ASR task.

\begin{table}[t]
\centering
\footnotesize
\caption{\textbf{Architecture-adapted ICAP-Gate configurations.} The table reports each selected pathway $(\ell_m)$, normalized depth, attenuation scale $(\lambda_m)$, and full-split confirmation scope.}
\label{tab:pathway-summary}
\setlength{\tabcolsep}{3.0pt}
\renewcommand{\arraystretch}{1.15}
\renewcommand{\tabularxcolumn}[1]{m{#1}}
\newcommand{\pathhead}[1]{\parbox[c][2.6\baselineskip][c]{\linewidth}{\centering\bfseries #1}}
\begin{tabularx}{\textwidth}{>{\centering\arraybackslash}m{0.22\textwidth}>{\raggedright\arraybackslash}X>{\centering\arraybackslash}m{0.14\textwidth}>{\centering\arraybackslash}m{0.06\textwidth}>{\centering\arraybackslash}m{0.18\textwidth}}
\toprule
\rowcolor{TableHeader}
\pathhead{Model} & \pathhead{Pathway $\ell_m$} & \pathhead{Depth\\(block/total)} & \pathhead{$\lambda_m$} & \pathhead{Full-split\\confirmation} \\
\midrule
\textbf{Qwen2.5-Omni-7B} & \textbf{\texttt{audio\_tower.}\allowbreak\texttt{layers.25}} & 0.81 (26/32) & \textbf{0.25} & ARC/MMLU $\times$ FSD/SIB \\
\rowcolor{TableAlt}
\textbf{Qwen2.5-Omni-3B} & \textbf{\texttt{audio\_tower.}\allowbreak\texttt{layers.25}} & 0.81 (26/32) & \textbf{0.25} & ARC/MMLU $\times$ FSD/SIB \\
\textbf{Phi-4-MM} & \textbf{\texttt{audio\_embed.encoder.}\allowbreak\texttt{encoders.23}} & 1.00 (24/24) & \textbf{0.10} & ARC/MMLU $\times$ FSD/SIB \\
\rowcolor{TableAlt}
\textbf{Voxtral-Mini-3B} & \textbf{\texttt{audio\_tower.}\allowbreak\texttt{layers.30}} & 0.97 (31/32) & \textbf{0.25} & ARC/MMLU $\times$ FSD/SIB \\
\bottomrule
\end{tabularx}
\end{table}

\subsection{Pathway Control Preserves Evaluated ASR Capability}
\label{sec:audio-safety}

Selective control preserves the pathway for explicit audio-demand instructions while attenuating it for text-only instructions; Figure~\ref{fig:selective-control-summary}(a) and Table~\ref{tab:selectivity} test these two requirements.

Fixed pathway suppression increases WER across all four architectures, whereas ICAP-Gate matches ungated WER on the evaluated ASR task for every model (Figure~\ref{fig:selective-control-summary}(a); Table~\ref{tab:selectivity}(a)). This demonstrates the conditional utility of the localized late audio pathway: attenuating its influence is beneficial when audio is irrelevant, while preserving it supports the evaluated ASR task.

Table~\ref{tab:selectivity}(b) verifies the evaluated routing scope: instruction-scoped routing preserves explicit audio requests and avoids false positives on question-content hard negatives, unlike full-query matching. Routing on the task instruction therefore aligns pathway control with modality demand rather than incidental lexical content; the complete audit and ASR diagnostics are provided in Appendices~\ref{app:routing-audit} and \ref{app:asr-safety}.

\subsection{Transcript Audit Supports Task-Irrelevant Speech Construction}
\label{sec:sib-validity}

SIB tests irrelevant audio with coherent linguistic structure. All MMLU-SIB transcripts were screened for IDF overlap; none of the 28 manually reviewed candidates was potentially relevant or answer-bearing, supporting the task-irrelevant construction of this condition (Table~\ref{tab:sib-validity}). Full construction and audit details are reported in Appendix~\ref{app:sib-validity}.

\begin{wraptable}{r}{0.60\linewidth}
\centering
\scriptsize
\caption{\textbf{MMLU-SIB construction and transcript audit.} All 14,042 transcripts were recovered; 28 IDF-overlap candidates were manually reviewed, and none was potentially relevant or answer-bearing. Details are in Appendix~\ref{app:sib-validity}.}
\label{tab:sib-validity}
\setlength{\tabcolsep}{2.5pt}
\renewcommand{\arraystretch}{1.04}
\begin{tabular}{@{}>{\raggedright\arraybackslash}p{0.75\linewidth}r@{}}
\toprule
\cellcolor{TableHeader}\textbf{Audit item} & \cellcolor{TableHeader}\textbf{Result} \\
\midrule
MMLU-SIB pairs & 14,042 \\
\cellcolor{TableAlt}Unique source utterances & \cellcolor{TableAlt}2,694 \\
Speaker/chapter coverage & 40/97 \\
\cellcolor{TableAlt}Mean/max utterance reuse & \cellcolor{TableAlt}5.212/14 \\
Transcripts recovered & 14,042/14,042 \\
\cellcolor{TableAlt}IDF-overlap candidates & \cellcolor{TableAlt}28 (0.20\%) \\
Potentially relevant screened candidates & 0/28 \\
\cellcolor{TablePositive}Answer-bearing speech & \cellcolor{TablePositive}\textbf{0/28} \\
\bottomrule
\end{tabular}
\end{wraptable}

\section{Analysis and Implications}
\label{sec:analysis}

\subsection{From Robustness to Selective Modality Influence}
\label{sec:selective-modality-analysis}

Paired Influence Rate, Answer Flip, and transition measures show why aggregate Accuracy is insufficient: irrelevant audio can change individual decisions while leaving the net score nearly unchanged. In these evaluations, reasoning stability is therefore characterized through \emph{modality influence} alongside final task score. The late pathway is conditionally useful: attenuating it reduces drift when audio is irrelevant, whereas preserving it maintains evaluated ASR capability. ICAP-Gate treats modality influence as a task-conditioned control property, yielding the principle that modality availability should not imply modality influence.

\subsection{Cross-Model Evidence Supports a Shared Computational Role}
\label{sec:cross-architecture-generalization}

\begin{figure}[t]
\centering
\includegraphics[width=0.8\textwidth]{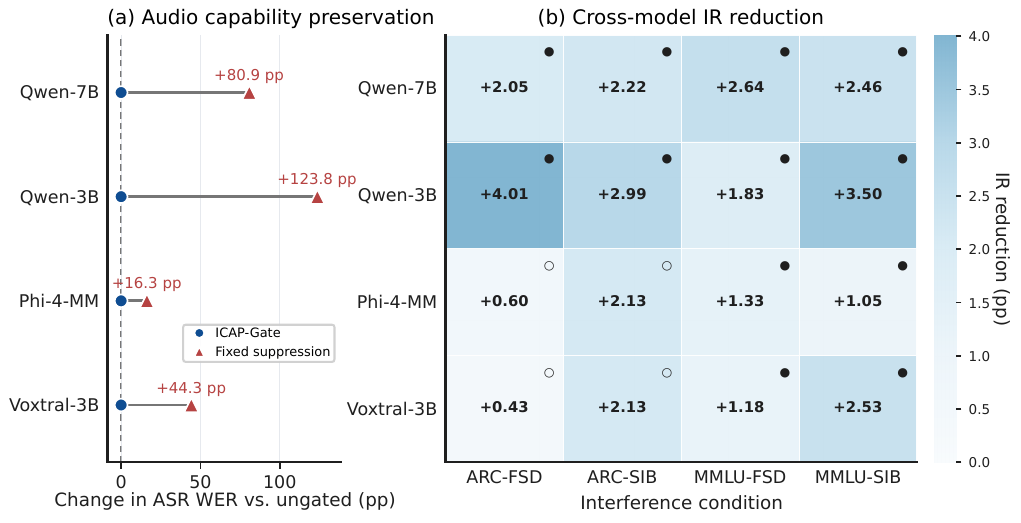}
\caption{\textbf{Selective control preserves evaluated ASR while reducing reasoning drift across architectures.} \textbf{(a)} Change in ASR WER relative to ungated inference on 400 LibriSpeech \texttt{dev-clean} utterances per model. Blue circles denote ICAP-Gate, red triangles fixed suppression, and the dashed line the ungated reference. \textbf{(b)} Full-split IR reduction (ungated minus ICAP-Gate); positive values indicate lower drift under ICAP-Gate. In Panel~(b), filled and hollow circles indicate whether the paired 95\% confidence interval excludes or overlaps zero, respectively. Phi-4-MM uses the canonical $s=0.10$ condition.}
\label{fig:selective-control-summary}
\end{figure}

Figure~\ref{fig:selective-control-summary}(b) and Table~\ref{tab:pathway-summary} show that the cross-model evidence supports a shared computational role rather than a shared layer identity. Qwen2.5-Omni, Phi-4-MM, and Voxtral-Mini-3B use distinct audio stacks, yet actionable sites recur in each model's late audio pathway. With model-specific pathway and attenuation parameters, the point estimates for Influence Rate and Answer Flip are lower than under ungated inference in all 16 full-split model--condition evaluations, supporting a shared task-conditioned rule with architecture-specific realizations. The corresponding model-specific analyses are summarized in Appendices~\ref{app:phi-validation} and \ref{app:voxtral-validation}. Within the evaluated environmental-sound and natural-speech conditions, modality demand is specified at the instruction level and the shared rule uses architecture-adapted pathway and attenuation parameters.

\begin{table}[t]
\centering
\caption{\textbf{Selectivity of ICAP-Gate.}
\textbf{(a)} ASR WER (\%, $\downarrow$) and recovery from fixed attenuation.
\textbf{(b)} Instruction-scoped routing on selected categories from the $n=240$ hard-case audit; FP denotes false positives.}
\label{tab:selectivity}

\begin{minipage}[t]{0.535\linewidth}
\centering
\textbf{(a) Audio-relevant preservation}\par
\footnotesize
\setlength{\tabcolsep}{2.2pt}
\renewcommand{\arraystretch}{1.04}

\begin{tabular}{lcccc}
\toprule
\rowcolor{black!10}
\textbf{Model} &
\textbf{Ungated} &
\textbf{Fixed} &
\textbf{ICAP} &
\textbf{Recovery$\uparrow$} \\
\midrule
Qwen2.5-Omni-7B
& 17.74 & 98.60
& \textbf{17.74}
& \textbf{80.86} \\
Qwen2.5-Omni-3B
& 60.82 & 184.58
& \textbf{60.82}
& \textbf{123.76} \\
Phi-4-MM (5.6B)
& 2.45 & 18.73
& \textbf{2.45}
& \textbf{16.28} \\
Voxtral-Mini-3B
& 16.72 & 61.05
& \textbf{16.72}
& \textbf{44.33} \\
\bottomrule
\end{tabular}
\end{minipage}
\hfill
\begin{minipage}[t]{0.445\linewidth}
\centering
\textbf{(b) Instruction-scoped routing}\par
\footnotesize
\setlength{\tabcolsep}{1.5pt}
\renewcommand{\arraystretch}{1.04}

\begin{tabular}{lccc}
\toprule
\rowcolor{black!10}
\textbf{Category} &
\textbf{Target} &
\textbf{Instr.} &
\textbf{Full query} \\
\midrule
Explicit audio req.
& High
& \textbf{40/40}
& 40/40 \\
Ordinary text instr.
& Low
& \textbf{40/40}
& 40/40 \\
\shortstack[l]{Question-content\\hard negatives}
& Low
& \textbf{0/40 FP}
& 40/40 FP \\
\bottomrule
\end{tabular}
\end{minipage}
\end{table}

\section{Conclusion}
\label{sec:conclusion}
We identify intervention-sensitive late audio pathways as a source of irrelevant-audio drift in large audio-language models. ICAP-Gate conditionally modulates each model's pathway according to task demand, reducing paired reasoning drift while matching ungated performance on the evaluated ASR task. Across four heterogeneous LALMs, two reasoning benchmarks, and environmental-sound and natural-speech interference, its point estimates for Influence Rate and Answer Flip are lower than under ungated inference in all 16 full-split model--condition evaluations. It has lower paired-drift point estimates than mitigation prompting across the four evaluated settings and provides competitive stability relative to eight-sample Self-Consistency while using one generation per query. These findings establish \emph{selective modality influence control} as a mechanism-guided principle for robust multimodal reasoning.

\subsection*{AI use statement}

In this work, generative AI tools were used only to improve the grammar, clarity, and readability of the manuscript. They were not used to generate data, formulate mathematical claims, design or implement the method, design or analyze experiments, process datasets, or interpret experimental results. All AI-assisted language edits were reviewed and revised by the authors. The authors take full responsibility for the final text, claims, analyses, code, figures, and other artifacts in this work.




\subsection*{Reproducibility statement}

The main paper specifies the paired evaluation protocol, reasoning-drift metrics, intervention rule, ICAP-Gate formulation, model-specific pathways, and full-split experimental settings. The appendices provide the online inference procedure, decoding and response settings, dataset construction and transcript audits, statistical checks, mechanistic localization analyses, matched mitigation comparisons, routing audits, and ASR safety evaluations.
We welcome independent reproduction and follow-up work.



\bibliography{iclr2027_conference}
\bibliographystyle{iclr2027_conference}

\appendix
\section*{Appendix}
\renewcommand{\arraystretch}{1.08}

\section{Reproducibility and Evaluation Protocol}
\label{app:reproducibility}

This appendix provides the implementation and validation details for the evaluation protocol defined in Section~\ref{sec:methodology}, including dataset configurations, paired-record requirements, answer extraction, uncertainty estimation, and integrity checks.

\subsection{Decoding and Response Settings}
\label{app:decoding-settings}

Qwen2.5-Omni runs use deterministic decoding. Voxtral-Mini-3B uses nucleus sampling with temperature $T=0.2$, $p=0.95$, and \texttt{max\_new\_tokens}=1024. The same model-specific decoding configuration is used for clean, ungated-audio, and ICAP-Gate conditions. Full-MMLU outputs follow the first-response truncation protocol before answer extraction.

\subsection{Evaluation Scope}
\label{app:evaluation-scope}

The evaluation spans four LALMs and the four full-split conditions summarized in Table~\ref{tab:app-evaluation-matrix}. Mechanistic localization is most detailed for Qwen2.5-Omni-7B, while all four architectures contribute full-split behavioral evidence for architecture-adapted pathway control across model scale and architecture.

The audio input is not required by the answer specification of the ARC and MMLU reasoning tasks. The clean condition therefore removes the audio input, whereas the ungated condition provides the audio input to the model through its native multimodal pathway. The gated condition uses the same audio input as the ungated condition but applies a pathway intervention selected by the evaluated gating policy.

\subsection{Paired Inference Conditions}
\label{app:paired-conditions}

The clean, ungated, and gated conditions follow Equation~\ref{eq:prediction-conditions}. Records are aligned by sample identifier, source index, question, choices, and gold answer before any paired metric is computed. Each output retains the generated response, extracted answer, gate decision, applied scale, matched modules, and available run metadata. Clean inference defines audio-induced change, while correctness is evaluated against the gold answer.

\subsection{Dataset Construction}
\label{app:dataset-construction}

\paragraph{FSD interference.}
FSD conditions pair the reasoning examples with unrelated environmental recordings from FSD50K. Aligned clean, ungated, and gated conditions preserve the question set across interventions.

\paragraph{SIB interference.}
SIB uses five-second, 16-kHz mono LibriSpeech dev-clean segments paired independently of question content. ARC-SIB samples without replacement, whereas MMLU-SIB samples with replacement. Appendix~\ref{app:sib-validity} reports construction, reuse, and transcript-level validity checks.

\paragraph{ASR side-effect evaluation.}
We evaluate 400 LibriSpeech dev-clean utterances under ungated inference, fixed suppression at $s=0.25$, and ICAP-Gate. WER against the reference transcription measures preservation on an explicitly audio-dependent task.

Each model uses an architecture-adapted late audio pathway and a model-specific attenuation scale; the selected configurations are summarized in the main-text pathway table.

\begin{table}[t]
\centering
\footnotesize
\setlength{\tabcolsep}{10pt}
\caption{Evaluation matrix used in the main full-split experiments. The table records the task, interference source, and number of paired examples.}
\label{tab:app-evaluation-matrix}
\begin{tabular}{llll}
\toprule
\rowcolor{TableHeader}
\textbf{Task} & \textbf{Reasoning dataset} & \textbf{Interference source} & \textbf{Examples} \\
\midrule
ARC-FSD & ARC-Challenge & FSD50K environmental sound & \textbf{1,172} \\
ARC-SIB & ARC-Challenge & LibriSpeech dev-clean speech & \textbf{1,172} \\
\rowcolor{TableAlt}
MMLU-FSD & MMLU & FSD50K environmental sound & \textbf{14,042} \\
\rowcolor{TableAlt}
MMLU-SIB & MMLU & LibriSpeech dev-clean speech & \textbf{14,042} \\
\bottomrule
\end{tabular}
\end{table}

\subsection{Answer Extraction and Response Handling}
\label{app:answer-extraction}

The reasoning tasks require the model to provide a multiple-choice answer. The inference protocol uses the task-specific prompt format and records the complete generated response. The evaluated answer is extracted from the response using the corresponding answer parser, which maps the first matching answer marker to a normalized choice or returns an invalid parse for validation. Responses follow the configured stop strings and the paper's primary response truncation mode. A stricter first-answer-line analysis applies the same parser after truncating each response to its first answer line; it is used as a robustness check rather than as a second metric definition. The parser implementation, stop strings, and extraction settings are included in the anonymous supplementary code archive.

The primary evaluation preserves the first-human response boundary defined by the run configuration. The post-hoc first-answer-line analysis recomputes the same metrics after truncating each response to its first answer line. This check separates answer-decision changes from response-continuation effects.

The clean, ungated, and gated records are required to have matching sample identity and compatible task metadata before paired metrics are computed. Validation scripts report and exclude records with mismatched source indices, missing answers, or incompatible task fields.

\subsection{Metrics}
\label{app:metrics}

Accuracy, IR, Answer Flip, NetC2W, and NetAnswer follow Equations~\ref{eq:accuracy}--\ref{eq:net-answer}. The implementation computes every metric from the same aligned records. IR operates on correctness states, while Answer Flip operates on answer identity; NetC2W summarizes correctness repair minus damage, and NetAnswer summarizes clean-answer restoration minus break. This distinction is retained in all tables and case extractions.

\subsection{Instruction-Conditioned Routing}
\label{app-routing-protocol}

ICAP-Gate applies the model-specific low scale to text-reasoning instructions and $s=1.0$ to explicit ASR or transcription requests. Each record stores the decision scope, matched cues, scales, and modules. The production rule parses only the instruction, preventing question-content terms from triggering the open pathway. Appendix~\ref{app:routing-audit} evaluates this decision rule on 240 manually curated hard cases.

\subsection{Online ICAP-Gate Inference Procedure}
\label{app:icap-inference}

The online procedure applies the calibrated architecture-specific control point without a clean-reference pass or a second generation. The router reads only the task instruction: explicit audio-demand instructions keep the pathway open, while other instructions use the model-specific attenuation scale. The pathway hook is temporary and is removed immediately after generation, so the same model remains available for subsequent requests.

Algorithm~\ref{alg:icap-gate} summarizes this single-generation online procedure.

\begin{algorithm}[t]
    \caption{Online ICAP-Gate inference}
    \label{alg:icap-gate}
    \begin{algorithmic}[1]
        \Require Model $f_m$, query $Q$, audio $A$, calibrated $(\ell_m,\lambda_m)$
        \Ensure Generated response $\widehat{y}$
        \State $I \gets \Call{ExtractInstruction}{Q}$
        \State $r \gets \Call{RequiresAudio}{I}$
        \State $g \gets 1$ if $r=1$, else $\lambda_m$
        \State $H \gets \Call{AttachTemporaryScale}{f_m,\ell_m,g}$
        \State $\widehat{y} \gets \Call{GenerateOnce}{f_m,Q,A}$
        \State \Call{Remove}{$H$}
        \State \Return $\widehat{y}$
    \end{algorithmic}
\end{algorithm}

Here $\Call{RequiresAudio}$ implements the deterministic instruction-scope cue rule described above. The scale is applied to the selected pathway representation, $\widetilde{\mathbf z}_{\ell_m}=g\mathbf z_{\ell_m}$, only during the current generation. Thus $g=1$ preserves the audio pathway for explicit audio-demand instructions, whereas $g=\lambda_m$ attenuates its downstream influence for text-reasoning instructions whose answers do not require audio.

\subsection{Statistical Analysis}
\label{app:statistical-analysis}

All primary reasoning comparisons are paired at the example level. Accuracy, Influence Rate, Answer Flip, and transition counts are computed from aligned clean, ungated, and gated records. Where raw paired outputs are available, uncertainty is estimated with paired bootstrap resampling over examples. Sign tests are used for selected transition analyses to test whether repair and damage directions are symmetric among discordant paired cases.

The reported confidence intervals quantify uncertainty over the evaluated examples. They do not quantify variation across model checkpoints, inference backends, or independent stochastic decoding seeds. This distinction is especially relevant for external models with task- and scale-dependent behavior.

The routing audit uses nominal Wilson intervals for its binary scope-level summaries. The audit contains matched scenario families and repeated templates, so these intervals describe the controlled audit and should not be interpreted as population-generalization intervals.

\paragraph{Self-Consistency audit.}
Across 8,688 clean and interference records, 328 ties occurred ($3.78\%$). The implementation deterministically selected the tied option appearing first in generation order. Unparseable generations, all-abstention cases, and final prediction extraction failures were all zero.

\subsection{Paired Bootstrap Input Audit}
\label{app:bootstrap-input-audit}

The paired bootstrap analysis used 20,000 percentile-bootstrap replicates with seed 0 and sample-aligned records. Query, sample-identifier, gold-answer, and MMLU-subject mismatches were zero, and the largest point-estimate discrepancy after recomputation was 0.009 percentage points.

For MMLU, subject-stratified resampling was used as a robustness check while preserving the observed subject sample sizes; its directional conclusions agree with the IID bootstrap.

These checks establish the alignment and numerical reproducibility of the reported paired estimates without listing the underlying artifact inventory.

\subsection{Protocol Scope}
\label{app:protocol-boundaries}

The protocol covers multiple-choice ARC and MMLU reasoning under FSD and randomly paired speech, with LibriSpeech ASR as the audio-relevant safety task. Paths and scales are model-specific. The SIB scope comprises random five-second speech segments; semantically similar, conversational, multi-speaker, and adversarial speech remain future evaluation settings.

\section{Baseline Reproduction}
\label{app:baseline-reproduction}

This appendix reports the paired clean and ungated baselines used to quantify native audio-induced instability before pathway intervention.

\begin{table}[t]
\centering
\small
\caption{Clean and ungated baseline results for the sixteen full-split model--task combinations. Accuracy is reported as clean/ungated, while IR and Answer Flip are measured relative to clean inference. All values are percentages.}
\label{tab:app-baseline-full}
\setlength{\tabcolsep}{4.5pt}
\begin{tabular}{llrrrr}
\toprule
\rowcolor{TableHeader}
\textbf{Model} & \textbf{Condition} & \textbf{Clean Acc.} & \textbf{Ungated Acc.} & \textbf{Ungated IR} & \textbf{Ungated Flip} \\
\midrule
\rowcolor{TableGroup}
Qwen2.5-Omni-7B & ARC-FSD & 84.56 & 83.53 & 8.19 & 9.04 \\
Qwen2.5-Omni-7B & ARC-SIB & 84.56 & 82.85 & 9.39 & 11.09 \\
\rowcolor{TableAlt}
Qwen2.5-Omni-7B & MMLU-FSD & 67.42 & 67.58 & 11.80 & 16.06 \\
\rowcolor{TableAlt}
Qwen2.5-Omni-7B & MMLU-SIB & 67.42 & 66.00 & 13.65 & 18.89 \\
\midrule
\rowcolor{TableGroup}
Qwen2.5-Omni-3B & ARC-FSD & 77.39 & 77.56 & 11.77 & 13.82 \\
Qwen2.5-Omni-3B & ARC-SIB & 77.39 & 76.54 & 13.82 & 16.13 \\
\rowcolor{TableAlt}
Qwen2.5-Omni-3B & MMLU-FSD & 60.36 & 59.76 & 15.24 & 21.61 \\
\rowcolor{TableAlt}
Qwen2.5-Omni-3B & MMLU-SIB & 60.36 & 58.97 & 17.79 & 25.17 \\
\midrule
\rowcolor{TableGroup}
Phi-4-MM & ARC-FSD & 81.66 & 75.51 & 22.53 & 25.00 \\
Phi-4-MM & ARC-SIB & 81.66 & 73.21 & 23.46 & 26.54 \\
\rowcolor{TableAlt}
Phi-4-MM & MMLU-FSD & 63.99 & 57.28 & 29.38 & 39.56 \\
\rowcolor{TableAlt}
Phi-4-MM & MMLU-SIB & 63.99 & 57.25 & 30.00 & 40.51 \\
\midrule
\rowcolor{TableGroup}
Voxtral-Mini-3B & ARC-FSD & 75.77 & 73.38 & 21.33 & 26.02 \\
Voxtral-Mini-3B & ARC-SIB & 75.77 & 69.88 & 25.17 & 30.20 \\
\rowcolor{TableAlt}
Voxtral-Mini-3B & MMLU-FSD & 57.79 & 58.00 & 27.44 & 37.53 \\
\rowcolor{TableAlt}
Voxtral-Mini-3B & MMLU-SIB & 57.79 & 56.62 & 28.41 & 39.52 \\
\bottomrule
\end{tabular}
\end{table}

\begin{table}[t]
\centering
\scriptsize
\setlength{\tabcolsep}{5pt}
\caption{Representative Qwen transition examples under task-irrelevant audio. The examples illustrate paired transition categories used by the analysis and are not an additional statistical evaluation.}
\label{tab:app-representative-cases}
\resizebox{\linewidth}{!}{%
\begin{tabular}{lll rllll}
\toprule
\rowcolor{TableHeader}
\textbf{Model} & \textbf{Condition} & \textbf{Subject} & \textbf{Index} & \textbf{Gold} & \textbf{Clean} & \textbf{Ungated} & \textbf{ICAP-Gate} \\
\midrule
Qwen2.5-Omni-7B & MMLU-FSD & abstract\_algebra & 0 & B & B & A & B \\
\rowcolor{TableAlt}
Qwen2.5-Omni-7B & MMLU-SIB & anatomy & 100 & A & A & D & A \\
Qwen2.5-Omni-3B & MMLU-FSD & clinical\_knowledge & 502 & C & C & A & C \\
\rowcolor{TableAlt}
Qwen2.5-Omni-3B & MMLU-SIB & astronomy & 243 & D & D & B & D \\
Qwen2.5-Omni-7B & MMLU-FSD & college\_physics & 1373 & B & A & D & A \\
\rowcolor{TableAlt}
Qwen2.5-Omni-3B & MMLU-SIB & college\_medicine & 1204 & A & A & B & A \\
\bottomrule
\end{tabular}
}
\end{table}
\subsection{Relation to the Original Phenomenon Study}
\label{app:baseline-relation}

\begin{table}[t]
\centering
\small
\caption{Self-Consistency tie audit for the matched mitigation comparisons. Tie rates are computed over the corresponding clean or interference records.}
\label{tab:app-scs-ties}
\setlength{\tabcolsep}{5pt}
\begin{tabular}{lrrrr}
\toprule
\rowcolor{TableHeader}
\textbf{Setting} & \textbf{Clean ties} & \textbf{Interference ties} & \textbf{Clean tie rate} & \textbf{Interference tie rate} \\
\midrule
ARC-FSD & 25 & 26 & 2.13\% & 2.22\% \\
ARC-SIB & 25 & 23 & 2.13\% & 1.96\% \\
\rowcolor{TableAlt}
MMLU-FSD ($n=1{,}000$) & 57 & 59 & 5.70\% & 5.90\% \\
\rowcolor{TableAlt}
MMLU-SIB ($n=1{,}000$) & 57 & 56 & 5.70\% & 5.60\% \\
\bottomrule
\end{tabular}
\end{table}

\subsection{Clean and Ungated Full-Split Comparisons}
\label{app:baseline-full-split}

Table~\ref{tab:app-baseline-full} reports clean Accuracy and ungated Accuracy, IR, and Answer Flip for all sixteen full-split model--task combinations.

Audio changes decisions across every model and both interference families, with larger drift in several Phi-4-MM and Voxtral conditions. This heterogeneity motivates model-specific pathway selection.

\subsection{Representative Transition Examples}
\label{app:representative-transitions}

Table~\ref{tab:app-representative-cases} lists representative Qwen transitions that make the paired repair and restoration categories concrete. The examples are selected by transition category and are qualitative rather than an additional statistical evaluation.

The baseline protocol follows the same phenomenon-level question as the original irrelevant-audio study: can an audio-language model change its text answer when the audio is not required? The present evaluation extends that comparison in three ways. First, it uses paired IR and Answer Flip measurements in addition to aggregate Accuracy. Second, it evaluates both structured environmental sound and random speech interference. Third, it preserves sample-aligned records for later pathway intervention and repair--damage analysis.

The baseline results establish the behavioral phenomenon and define the reference condition for ICAP-Gate. They motivate sample-level stability analysis and pathway intervention across models.

\subsection{Matched Mitigation Comparison}
\label{app:matched-mitigation}

The matched mitigation analysis compares the Prompting, Self-Consistency, and ICAP-Gate inference paths against each method's own clean reference. Table~\ref{tab:app-mitigation-comparison} reports the complete matched-method comparison. Prompting and ICAP-Gate use the full ARC and MMLU splits; Self-Consistency uses the full ARC splits and fixed matched MMLU subsets with $n=1{,}000$. Prompting is applied once before generation, whereas Self-Consistency generates eight responses under the same task prompt with temperature $0.5$ and $p=1.0$, then selects the modal extracted answer; ties are resolved by the first generation in order, as audited in Table~\ref{tab:app-scs-ties}. The exact prompt template and generation arguments are included in the anonymous supplementary code archive. All records passed source-ID alignment checks, and final extraction failures, all-abstention cases, and unparseable Self-Consistency generations were zero.

The paired differences show that Prompting has higher drift point estimates than ICAP-Gate in all four settings, with confidence intervals excluding zero on ARC-SIB and both full MMLU comparisons. Self-Consistency has lower drift point estimates in all four matched comparisons, with a statistically resolved advantage on ARC-SIB and no clear difference in the other three comparisons. These results support the paper-facing distinction between input-level guidance, output-level aggregation, and mechanism-guided pathway control.

\begin{table}[t]
\centering
\scriptsize
\caption{Matched mitigation comparison. Panel (a) reports method-relative clean/interference outcomes; Panel (b) reports comparator-minus-ICAP-Gate paired differences in IR and Answer Flip with 95\% paired percentile-bootstrap confidence intervals from 20,000 resamples (seed 0). Positive differences in Panel (b) indicate lower drift under ICAP-Gate.}
\label{tab:app-mitigation-comparison}

\setlength{\tabcolsep}{3pt}


{\small\bfseries (a) Method-relative outcomes.\par}

{\renewcommand{\arraystretch}{1.04}
\resizebox{\linewidth}{!}{%
\begin{tabular}{lllrrrrrl}
\toprule
\rowcolor{TableHeader}
\textbf{Setting}
& \textbf{Method}
& \textbf{Sample}
& \textbf{\shortstack{Clean\\Acc.}}
& \textbf{\shortstack{Interference\\Acc.}}
& \textbf{\shortstack{$\Delta$\\Acc.}}
& \textbf{IR}
& \textbf{\shortstack{Answer\\Flip}}
& \textbf{\shortstack{Repair\\/Damage\\/Net}}
\\
\midrule

\rowcolor{TablePositive}
ARC-FSD
& \textbf{ICAP-Gate}
& Full
& 84.56\%
& 84.90\%
& +0.34
& 6.14\%
& 7.00\%
& 38/34/+4
\\

ARC-FSD
& Prompting
& Full
& 84.98\%
& 83.96\%
& -1.02
& 7.68\%
& 8.62\%
& 39/51/-12
\\

ARC-FSD
& Self-Consistency
& Full
& 87.37\%
& 87.80\%
& +0.43
& 4.52\%
& 5.20\%
& 29/24/+5
\\

\midrule

\rowcolor{TablePositive}
ARC-SIB
& \textbf{ICAP-Gate}
& Full
& 84.56\%
& 85.07\%
& +0.51
& 7.17\%
& 8.28\%
& 45/39/+6
\\

ARC-SIB
& Prompting
& Full
& 84.98\%
& 81.31\%
& -3.67
& 10.32\%
& 12.12\%
& 39/82/-43
\\

ARC-SIB
& Self-Consistency
& Full
& 87.37\%
& 87.12\%
& -0.26
& 4.18\%
& 5.38\%
& 23/26/-3
\\

\midrule

\rowcolor{TablePositive}
MMLU-FSD
& \textbf{ICAP-Gate}
& Full
& 67.42\%
& 67.55\%
& +0.13
& 9.16\%
& 12.60\%
& 652/634/+18
\\

MMLU-FSD
& Prompting
& Full
& 67.28\%
& 66.86\%
& -0.41
& 12.82\%
& 17.32\%
& 871/929/-58
\\

\rowcolor{TablePositive}
MMLU-FSD
& \textbf{ICAP-Gate}
& Matched $n=1{,}000$
& 66.90\%
& 68.30\%
& +1.40
& 9.60\%
& 12.80\%
& 55/41/+14
\\

MMLU-FSD
& Self-Consistency
& Matched $n=1{,}000$
& 69.10\%
& 68.30\%
& -0.80
& 7.80\%
& 11.70\%
& 35/43/-8
\\

\midrule

\rowcolor{TablePositive}
MMLU-SIB
& \textbf{ICAP-Gate}
& Full
& 67.42\%
& 67.68\%
& +0.26
& 11.19\%
& 15.58\%
& 804/767/+37
\\

MMLU-SIB
& Prompting
& Full
& 67.28\%
& 60.55\%
& -6.72
& 20.05\%
& 27.35\%
& 936/1880/-944
\\

\rowcolor{TablePositive}
MMLU-SIB
& \textbf{ICAP-Gate}
& Matched $n=1{,}000$
& 66.90\%
& 67.80\%
& +0.90
& 10.70\%
& 15.90\%
& 58/49/+9
\\

MMLU-SIB
& Self-Consistency
& Matched $n=1{,}000$
& 69.10\%
& 67.30\%
& -1.80
& 10.60\%
& 14.70\%
& 44/62/-18
\\

\bottomrule
\end{tabular}%
}}

\smallskip


{\small\bfseries (b) Paired differences relative to ICAP-Gate.\par}

{\renewcommand{\arraystretch}{1.20}
\resizebox{\linewidth}{!}{%
\begin{tabular}{llrrrr}
\toprule
\rowcolor{TableHeader}
\textbf{Setting}
& \textbf{Comparison}
& \textbf{$\Delta$IR}
& \textbf{\shortstack{95\%\\CI}}
& \textbf{$\Delta$Flip}
& \textbf{\shortstack{95\%\\CI}}
\\
\midrule

ARC-FSD
& Prompt--ICAP
& +1.54
& $[-0.17,+3.24]$
& +1.62
& $[-0.17,+3.41]$
\\

ARC-FSD
& SCS--ICAP
& -1.62
& $[-3.33,+0.09]$
& -1.79
& $[-3.58,0.00]$
\\

\rowcolor{TablePositive}
ARC-SIB
& Prompt--ICAP
& +3.16
& $[+1.19,+5.12]$
& +3.84
& $[+1.79,+5.89]$
\\

ARC-SIB
& SCS--ICAP
& -2.99
& $[-4.78,-1.28]$
& -2.90
& $[-4.69,-1.11]$
\\

\rowcolor{TablePositive}
MMLU-FSD
& Prompt--ICAP, full
& +3.66
& $[+3.04,+4.29]$
& +4.72
& $[+4.04,+5.41]$
\\

MMLU-FSD
& SCS--ICAP, $n=1{,}000$
& -1.80
& $[-4.10,+0.40]$
& -1.10
& $[-3.60,+1.50]$
\\

\rowcolor{TablePositive}
MMLU-SIB
& Prompt--ICAP, full
& +8.87
& $[+8.10,+9.63]$
& +11.76
& $[+10.95,+12.58]$
\\

MMLU-SIB
& SCS--ICAP, $n=1{,}000$
& -0.10
& $[-2.40,+2.20]$
& -1.20
& $[-3.70,+1.40]$
\\

\bottomrule
\end{tabular}%
}}

\end{table}

\section{Qwen Mechanistic Localization}
\label{app:qwen-mechanistic-localization}

This appendix summarizes the intervention evidence supporting the Qwen2.5-Omni late audio pathway and its confirmation on the locked full-split conditions.

\subsection{Diagnostic Traces and the Localization Question}
\label{app:qwen-diagnostic-traces}

We compare clean text-only inputs with paired audio conditions while recording intermediate representations and answer-token logits. Teacher-forced traces and progressive condition comparisons show that the audio condition changes both internal trajectories and the relative logits of answer options. These traces establish that the behavioral effect is accompanied by multimodal computation changes before the final extracted answer. They do not, by themselves, identify a unique intervention site.

Direct residual-state patching provides a complementary intervention test. Patching a single residual state does not consistently reproduce the full behavioral effect, indicating that the relevant disturbance is distributed beyond one isolated hidden-state substitution. We therefore use module-level ablation and output scaling to identify an intervention that is both behaviorally effective and operationally stable. The selected gate is thus a module-level control point supported by intervention evidence.

\subsection{Offline Pathway Calibration}
\label{app:qwen-offline-calibration}

The Qwen pathway and attenuation scale were calibrated offline on the prescribed selection split and then evaluated on the locked full-split conditions. Candidate late audio modules and attenuation scales were compared with the same paired-drift and transition-recovery criteria used to define an actionable control point; the selected configuration was fixed before any full-split evaluation. No full-split result was used to choose the pathway or scale. This separation keeps calibration provenance distinct from the confirmatory Qwen estimates reported below; the external architecture sections describe their corresponding candidate-intervention validation.
\subsection{Activation Statistics and the Directional Interpretation}
\label{app:qwen-activation-statistics}

To test whether the interference is explained simply by larger audio activations, we compare activation RMS and the clean gold-answer logit margin for clean-correct-to-wrong (C2W) cases and control cases. Table~\ref{tab:app-qwen-activation-stats} shows that the activation RMS changes little between the two groups, whereas the gold-answer margin differs substantially. The corresponding ratios are close to one at both late25 and late30. This pattern is consistent with a directional or fusion-related change in the representation-to-logit map rather than a generic increase in activation magnitude. The table reports the exact aggregate values visualized in Figure~\ref{fig:activation-margin-contrast}.

\begin{table}[t]
\centering
\small
\caption{Qwen2.5-Omni FSD activation statistics for C2W and control cases. RMS is the activation root-mean-square at the indicated late audio-tower site; the margin is the clean gold-answer logit margin.}
\label{tab:app-qwen-activation-stats}
\begin{tabular}{lrrrrr}
\toprule
\rowcolor{TableHeader}
\textbf{Site} & \textbf{C2W RMS} & \textbf{Control RMS} & \textbf{RMS ratio} & \textbf{C2W margin} & \textbf{Control margin} \\
\midrule
late25 & 1.5023 & 1.4833 & \textbf{1.0128} & 0.2098 & 3.7383 \\
\rowcolor{TableAlt}
late30 & 1.7011 & 1.6731 & \textbf{1.0168} & 0.2098 & 3.7383 \\
\bottomrule
\end{tabular}
\end{table}

The activation statistics show why raw magnitude is insufficient as a selection criterion. We therefore evaluate pathway direction and answer-logit stability together.

\subsection{Mechanistic Interpretation and Scope}
\label{app:qwen-mechanistic-scope}

Taken together, the drift traces, intervention evidence, and activation statistics identify \texttt{audio\_tower.layers.25} as a stable actionable site for Qwen2.5-Omni under the evaluated protocol. Qwen2.5-Omni-7B provides higher-resolution localization evidence, while the full-split results across all four models provide behavioral confirmation for architecture-adapted control. Qwen2.5-Omni-3B tests same-family transfer, and Phi-4-MM and Voxtral-Mini-3B are evaluated with model-specific late pathways and suppression strengths.

The mechanistic conclusion is specific to the evaluated Qwen pathway: intervention evidence identifies late audio-tower computation as a behaviorally actionable site for irrelevant-audio drift, and selective attenuation at a stable late site reduces decision drift. This result motivates model-specific pathway selection and instruction-conditioned control.

\section{Phi-4-MM External Validation}
\label{app:phi-validation}

Phi-4-MM provides an external architecture check after the Qwen2.5-Omni mechanism study. Its audio stack differs from Qwen's, so the intervention targets the late speech/audio encoder path \texttt{audio\_embed.encoder.encoders.23}. This analysis tests transfer across architectures and characterizes task- and scale-dependent responses. The results concern this selected late pathway and are reported separately from the primary Qwen intervention.

\subsection{Full ARC and MMLU Results}
\label{app:phi-full-results}

Table~\ref{tab:app-phi-full} reports the four full-split reasoning conditions. The clean/ungated/gated Accuracy values are accompanied by IR, Answer Flip, and paired transition summaries. The evaluated full Phi configuration uses encoder layer 23 with low scale $s=0.10$. All full reasoning rows contain 1,172 ARC examples or 14,042 MMLU examples, with source-index alignment preserved within each comparison.

\begin{table}[t]
\centering
\scriptsize
\setlength{\tabcolsep}{4.3pt}
\caption{Phi-4-MM full-split external validation. Accuracy is clean/ungated/gated; IR and Answer Flip are ungated $\rightarrow$ gated. Net C2W and Net Answer are paired transition summaries.}
\label{tab:app-phi-full}
\resizebox{\linewidth}{!}{%
\begin{tabular}{lrrrrrrr}
\toprule
\rowcolor{TableHeader}
\textbf{Condition} & \textbf{Acc. C/U/G} & \textbf{IR U$\rightarrow$G} & \textbf{Flip U$\rightarrow$G} & \textbf{Net C2W} & \textbf{Net Answer} & \textbf{Interpretation} \\
\midrule
ARC-FSD & 81.66/75.51/75.77 & 22.53$\rightarrow$21.93 & 25.00$\rightarrow$24.66 & +5 & +4 & Small drift reduction \\
\rowcolor{TableAlt}
ARC-SIB & 81.66/73.21/75.51 & 23.46$\rightarrow$21.33 & 26.54$\rightarrow$24.66 & +26 & +22 & Positive speech transfer \\
MMLU-FSD & 63.99/57.28/58.98 & 29.38$\rightarrow$28.04 & 39.56$\rightarrow$38.04 & +213 & +214 & Best full Phi setting \\
\rowcolor{TableAlt}
MMLU-SIB & 63.99/57.25/57.51 & 30.00$\rightarrow$28.96 & 40.51$\rightarrow$38.73 & +92 & +249 & Near-neutral accuracy \\
\bottomrule
\end{tabular}
}
\end{table}

The Phi full-split results provide external behavioral evidence for architecture-adapted late-pathway control: IR and Answer Flip have lower point estimates than under ungated inference in all four conditions, while the transition summaries quantify the associated correctness changes.

\subsection{Scale-Dependent MMLU Response}
\label{app:phi-scale-boundary}

Table~\ref{tab:app-phi-mmlu-scale} reports the two evaluated Phi-4-MM full-split scales. The comparison shows architecture- and task-dependent scale behavior; the reported full-split results use the selected scale for each condition.

\begin{table}[t]
\centering
\small
\setlength{\tabcolsep}{4.5pt}
\caption{Phi-4-MM full MMLU scale comparison. All rows use encoder layer 23 and 14,042 examples. Net Overall denotes repair minus damage over correctness transitions; positive values favor gating.}
\label{tab:app-phi-mmlu-scale}
\begin{tabular}{lrrrrrr}
\toprule
\rowcolor{TableHeader}
\textbf{Dataset} & \textbf{Scale} & \textbf{Gated Acc.} & \textbf{Gated IR} & \textbf{Gated Flip} & \textbf{Net C2W} & \textbf{Net Overall / Net Answer} \\
\midrule
\rowcolor{TablePositive}
MMLU-FSD & \textbf{0.10} & 58.98 & 28.04 & 38.04 & +213 & +239 / +214 \\
MMLU-FSD & 0.25 & 58.21 & 28.66 & 38.83 & +116 & +131 / +102 \\
\rowcolor{TablePositive}
MMLU-SIB & \textbf{0.10} & 57.51 & 28.96 & 38.73 & +92 & +37 / +249 \\
\rowcolor{TableAlt}
MMLU-SIB & 0.25 & 56.38 & 29.43 & 39.74 & -21 & -122 / +108 \\
\bottomrule
\end{tabular}
\end{table}

The evaluated scales produce different transition profiles across the two MMLU conditions. This result supports selecting the attenuation scale together with the architecture-adapted pathway rather than treating a single scale as universal.

\subsection{Cross-Model Transition Significance}
\label{app:cross-model-significance}

Table~\ref{tab:app-cross-model-significance} reports two-sided exact sign-test $p$-values over discordant paired transitions. The columns retain the transition definitions used by each full-split analysis; they do not represent bootstrap intervals.

\begin{table}[t]
\centering
\small
\caption{Unified transition significance for Phi-4-MM and Voxtral-Mini-3B. Each value is a two-sided exact sign-test $p$-value over discordant paired cases within the listed model--condition.}
\label{tab:app-cross-model-significance}
\setlength{\tabcolsep}{4pt}
\begin{tabular}{lllrrr}
\toprule
\rowcolor{TableHeader}
\textbf{Model} & \textbf{Condition} & \textbf{Configuration} & \textbf{\shortstack{C2W\\repair/damage}} & \textbf{\shortstack{Overall\\repair/damage}} & \textbf{\shortstack{Answer\\restore/break}} \\
\midrule
\rowcolor{TablePositive}
Phi-4-MM & MMLU-FSD & $s=0.10$ & \textbf{$4.31\times10^{-7}$} & \textbf{$6.36\times10^{-6}$} & \textbf{$9.13\times10^{-5}$} \\
\rowcolor{TablePositive}
Phi-4-MM & MMLU-FSD & $s=0.25$ & \textbf{$0.00321$} & \textbf{$0.00866$} & \textbf{$0.0481$} \\
Phi-4-MM & MMLU-SIB & $s=0.10$ & \textbf{$0.0378$} & $0.515$ & \textbf{$1.35\times10^{-5}$} \\
Phi-4-MM & MMLU-SIB & $s=0.25$ & $0.642$ & $0.0235$ & $0.0527$ \\
\rowcolor{TableAlt}
Voxtral-Mini-3B & ARC-FSD & $s=0.25$ & $0.646$ & $0.250$ & $0.951$ \\
\rowcolor{TablePositive}
Voxtral-Mini-3B & ARC-SIB & $s=0.25$ & $0.0296$ & $0.0206$ & $0.0498$ \\
Voxtral-Mini-3B & MMLU-FSD & $s=0.25$ & $0.0217$ & $0.619$ & \textbf{$0.000392$} \\
\rowcolor{TablePositive}
Voxtral-Mini-3B & MMLU-SIB & $s=0.25$ & $3.54\times10^{-14}$ & $4.38\times10^{-7}$ & $2.54\times10^{-13}$ \\
\bottomrule
\end{tabular}
\end{table}

\section{Voxtral-Mini-3B Cross-Model Validation}
\label{app:voxtral-validation}

Voxtral-Mini-3B provides a second external architecture check with a distinct audio stack. The intervention targets \texttt{audio\_tower.layers.30} at low scale $s=0.25$. The full-split results test model-specific late-pathway control across the four evaluated conditions.

\subsection{Full-Split Transfer Results}
\label{app:voxtral-full-results}

Table~\ref{tab:app-voxtral-full} expands the Voxtral rows in the main table. ARC contains 1,172 examples and MMLU contains 14,042 examples. Accuracy is shown for clean, ungated-interference, and gated-interference conditions; IR and Answer Flip are reported as ungated $\rightarrow$ gated. The full-split rows report the model-specific stability pattern across environmental sound and speech interference; aggregate Accuracy and drift are interpreted as separate outcomes.

\begin{table}[t]
\centering
\scriptsize
\setlength{\tabcolsep}{4.2pt}
\caption{Voxtral-Mini-3B full-split transfer results. Accuracy is clean/ungated/gated; IR and Answer Flip are ungated $\rightarrow$ gated. Net C2W and Net Answer are paired transition summaries.}
\label{tab:app-voxtral-full}
\resizebox{\linewidth}{!}{%
\begin{tabular}{lrrrrrr}
\toprule
\rowcolor{TableHeader}
\textbf{Condition} & \textbf{Acc. C/U/G} & \textbf{IR U$\rightarrow$G} & \textbf{Flip U$\rightarrow$G} & \textbf{Net C2W} & \textbf{Net Answer} & \textbf{Interpretation} \\
\midrule
ARC-FSD & 75.77/73.38/71.76 & 21.33$\rightarrow$20.90 & 26.02$\rightarrow$25.85 & $-7$ & $+2$ & Accuracy trade-off \\
\rowcolor{TableAlt}
ARC-SIB & 75.77/69.88/73.38 & 25.17$\rightarrow$23.04 & 30.20$\rightarrow$27.22 & $+33$ & $+35$ & Positive speech transfer \\
MMLU-FSD & 57.79/58.00/58.21 & 27.44$\rightarrow$26.26 & 37.53$\rightarrow$36.00 & $+98$ & $+215$ & Positive stability transfer \\
\rowcolor{TableAlt}
MMLU-SIB & 57.79/56.62/58.76 & 28.41$\rightarrow$25.88 & 39.52$\rightarrow$36.28 & $+328$ & $+454$ & Strongest full transfer \\
\bottomrule
\end{tabular}
}
\end{table}

The full matrix shows model-specific stabilization across environmental sound and speech interference. The table is an aggregate full-split validation and is interpreted through drift reduction and transition counts rather than subject-level case inventories.

\section{Instruction-Routing Hard-Case Audit}
\label{app:routing-audit}

This appendix characterizes the operating range of the instruction-conditioned router independently of model generation. It asks whether the cue rule preserves the pathway for audio requests and avoids misleading lexical triggers in text-only instructions.

\subsection{Operating Rule}
\label{app:routing-protocol}

The audit contains 240 manually curated instruction cases from 40 matched scenario families, balanced between 120 audio-required and 120 audio-irrelevant cases. Label one preserves the audio pathway and label zero permits attenuation. Six categories contain 40 cases each: explicit requests, implicit dependence, cue-free paraphrases, negated keywords, question-content hard negatives, and ordinary text negatives.

The same cases are evaluated under two scopes. The deployed policy inspects the instruction before the first blank line; the full-query condition also exposes the question and answer choices, testing scope contamination. Both runs process all 240 cases.

\subsection{Routing Coverage}
\label{app:routing-scope-results}

Table~\ref{tab:app-routing-audit} summarizes the instruction-scoped routing audit over the evaluated categories. The table reports the operating range of the deployed rule and the effect of exposing question content to the router.

Instruction scope prevents question-content keywords from changing decisions on the tested hard negatives. The category rows distinguish explicit requests, ordinary text, keyword hard negatives, and implicit or cue-free dependence.

\begin{table}[t]
\centering
\scriptsize
\caption{Instruction-scoped routing audit over 240 manually curated cases. Panel (a) reports scope-level counts and nominal Wilson intervals; Panel (b) reports category-level TP/TN/FP/FN counts. The audit characterizes the deployed instruction-scoped rule and is not a general semantic audio-relevance benchmark.}
\label{tab:app-routing-audit}
\setlength{\tabcolsep}{3.5pt}

{\small\bfseries (a) Scope-level results.\par}
\resizebox{\linewidth}{!}{%
\begin{tabular}{lrrrrr rrrrr}
\toprule
\rowcolor{TableHeader}
\textbf{Scope} & \textbf{TP} & \textbf{TN} & \textbf{FP} & \textbf{FN} & \textbf{Accuracy (95\% CI)} & \textbf{Precision} & \textbf{Recall} & \textbf{FPR} & \textbf{FNR} & \textbf{High-scale rate} \\
\midrule
Instruction & 40 & 80 & 40 & 80 & 50.00\% [43.72, 56.28] & 50.00\% & 33.33\% & 33.33\% & 66.67\% & 33.33\% \\
\rowcolor{TableAlt}
Full query & 40 & 40 & 80 & 80 & 33.33\% [27.67, 39.52] & 33.33\% & 33.33\% & 66.67\% & 66.67\% & 50.00\% \\
\bottomrule
\end{tabular}
}

\vspace{3pt}
{\small\bfseries (b) Category-level counts.\par}
\resizebox{\linewidth}{!}{%
\begin{tabular}{lclll}
\toprule
\rowcolor{TableHeader}
\textbf{Category} & \textbf{Reference class} & \textbf{Instruction result} & \textbf{Full-query result} & \textbf{Interpretation} \\
\midrule
explicit\_positive & Positive & 40/0/0/0 & 40/0/0/0 & Explicit cue coverage is complete. \\
\rowcolor{TableAlt}
implicit\_positive & Positive & 0/0/0/40 & 0/0/0/40 & Implicit audio dependence is missed. \\
paraphrase\_positive & Positive & 0/0/0/40 & 0/0/0/40 & Cue-free paraphrases are missed. \\
\rowcolor{TableAlt}
instruction\_keyword\_negative & Negative & 0/0/40/0 & 0/0/40/0 & Negated audio keywords trigger high scale. \\
keyword\_hard\_negative & Negative & 0/40/0/0 & 0/0/40/0 & Full-query scope adds question-content false positives. \\
\rowcolor{TableAlt}
ordinary\_text\_negative & Negative & 0/40/0/0 & 0/40/0/0 & Ordinary text remains low scale. \\
\bottomrule
\end{tabular}
}
\end{table}

\subsection{Implication}
\label{app:routing-reproducibility}

The audit bounds the claim to instruction-scoped routing over the evaluated categories; it is not a general semantic audio-relevance evaluation.

\section{Audio-Relevant Safety and ASR Side Effects}
\label{app:asr-safety}

This appendix provides the complete LibriSpeech ASR side-effect evaluation for an explicitly audio-dependent task. Each model is compared with its own ungated WER to measure functional preservation.

\subsection{Protocol}
\label{app:asr-protocol}

We evaluate 400 utterances from LibriSpeech dev-clean under three aligned conditions: ungated inference, fixed suppression with low scale $s=0.25$, and ICAP-Gate. The fixed condition applies the selected model-specific intervention to every utterance. ICAP-Gate uses the same low scale for non-audio instructions but restores the selected pathway to $s=1.0$ when the instruction explicitly requests speech recognition or transcription. The selected paths are \texttt{audio\_tower.layers.25} for both Qwen2.5-Omni models, \texttt{audio\_embed.encoder.encoders.23} for Phi-4-MM, and \texttt{audio\_tower.layers.30} for Voxtral-Mini-3B.

Word Error Rate (WER) is computed against the reference transcription, with lower values indicating better recognition. Because WER includes insertion errors, it can exceed 100\% and should not be interpreted as the fraction of incorrectly transcribed utterances. All conditions use the same utterance set and the same answer-extraction/evaluation protocol.

\subsection{ASR Integration}
\label{app:phi-claim-scope}

Phi-4-MM contributes both a late-audio-encoder transfer result and an ASR safety result. Fixed suppression at encoder layer 23 increases WER, whereas ICAP-Gate routes explicit ASR instructions to the open pathway and matches ungated WER. Together with the full MMLU analyses, these results support instruction-conditioned preservation across an external architecture.

\subsection{Fixed Suppression and Instruction-Conditioned Preservation}
\label{app:asr-results}

Table~\ref{tab:app-asr-safety} shows the safety contrast: fixed suppression raises WER in every architecture, whereas ICAP-Gate matches ungated WER by routing every ASR instruction to the open pathway.

Fixed-gate WER increases range from 16.28 to 123.76 points. Absolute ungated WER varies with the model and inference pipeline, so the comparison is within model.

\begin{table}[t]
\centering
\scriptsize
\caption{ASR safety and uncertainty on 400 LibriSpeech dev-clean utterances per model. Panel (a) reports point estimates for all four models; Panel (b) reports paired bootstrap intervals for the model-condition pairs with matched utterance-level summaries. WER differences are measured relative to ungated inference.}
\label{tab:app-asr-safety}
\setlength{\tabcolsep}{4.5pt}

{\small\bfseries (a) Four-model point estimates.\par}
\resizebox{\linewidth}{!}{%
\begin{tabular}{lrrrrrrr}
\toprule
\rowcolor{TableHeader}
\textbf{Model} & \textbf{Selected path} & \textbf{Ungated WER} & \textbf{Fixed WER} & \textbf{ICAP-Gate WER} & \textbf{Fixed $\Delta$} & \textbf{ICAP $\Delta$} & \textbf{High-scale routing} \\
\midrule
Qwen2.5-Omni-7B & \texttt{layers.25} & 17.74 & \cellcolor{TableBoundary}98.60 & \cellcolor{TablePositive}17.74 & \cellcolor{TableBoundary}+80.86 & \cellcolor{TablePositive}0.00 & \cellcolor{TablePositive}400/400 \\
\rowcolor{TableAlt}
Qwen2.5-Omni-3B & \texttt{layers.25} & 60.82 & \cellcolor{TableBoundary}184.58 & \cellcolor{TablePositive}60.82 & \cellcolor{TableBoundary}+123.76 & \cellcolor{TablePositive}0.00 & \cellcolor{TablePositive}400/400 \\
Phi-4-MM & \texttt{encoders.23} & 2.45 & \cellcolor{TableBoundary}18.73 & \cellcolor{TablePositive}2.45 & \cellcolor{TableBoundary}+16.28 & \cellcolor{TablePositive}0.00 & \cellcolor{TablePositive}400/400 \\
\rowcolor{TableAlt}
Voxtral-Mini-3B & \texttt{layers.30} & 16.72 & \cellcolor{TableBoundary}61.05 & \cellcolor{TablePositive}16.72 & \cellcolor{TableBoundary}+44.33 & \cellcolor{TablePositive}0.00 & \cellcolor{TablePositive}400/400 \\
\bottomrule
\end{tabular}
}

\vspace{3pt}

{\small\bfseries (b) Paired bootstrap intervals.\par}
{\small
\setlength{\tabcolsep}{6pt}
\renewcommand{\arraystretch}{1.08}
\begin{tabular}{lrr}
\toprule
\rowcolor{TableHeader}
\textbf{Model} & \textbf{\shortstack{Fixed $\Delta$ WER\\(95\% CI)}} & \textbf{\shortstack{ICAP $\Delta$ WER\\(95\% CI)}} \\
\midrule
Qwen2.5-Omni-7B & +80.86 [73.72, 88.53] & 0.00 [0.00, 0.00] \\
Phi-4-MM & +16.28 [12.71, 20.23] & 0.00 [0.00, 0.00] \\
\bottomrule
\end{tabular}
}

\end{table}

\subsection{Paired Uncertainty and Routing Audit}
\label{app:asr-uncertainty}

For the two models with matched utterance-level summaries, Table~\ref{tab:app-asr-safety} reports paired WER intervals; the ICAP-Gate difference is exactly zero in these audited runs because ASR instructions use the open scale.

All 400 ASR instructions for each model are marked audio-relevant and assigned the high scale, directly linking the routing decision to functional preservation. Appendix~\ref{app:routing-audit} evaluates implicit and cue-free requests separately.

\subsection{ASR Preservation}
\label{app:asr-boundary}

Fixed suppression remains useful for causal intervention under known irrelevant audio, whereas ICAP-Gate adds the instruction-level control required for explicit ASR tasks. Broader audio-understanding safety remains outside this evaluation.

\section{Speech Interference Benchmark Construction and Validity}
\label{app:sib-validity}

This appendix documents SIB construction, source coverage, reuse, and transcript-level validity. Each text question is paired with a natural speech segment selected independently of the question and answer.

\subsection{Source Corpus}
\label{app:sib-source-corpus}

The speech pool is LibriSpeech dev-clean with 40 speakers, 97 chapters, and approximately 5.39 hours of source speech. All retained SIB clips are processed as 16-kHz mono segments with a fixed five-second duration.

\subsection{Pairing and Reuse Statistics}
\label{app:sib-pairing-statistics}

ARC-SIB contains 1,172 ARC-Challenge text--speech pairs sampled without replacement, so every paired utterance is unique within ARC-SIB. MMLU-SIB contains 14,042 MMLU text--speech pairs sampled with replacement because the question set is larger than the available speech pool. MMLU-SIB uses 2,694 unique utterances, with mean reuse 5.212, median reuse 5, 95th-percentile reuse 9, and maximum reuse 14. Table~\ref{tab:app-sib-construction} reports the complete paired-dataset statistics.

\begin{table}[t]
\centering
\small
\caption{SIB source coverage and pairing statistics. Source rows describe the common LibriSpeech pool; pairing rows describe the two reasoning conditions. Processed duration counts paired clips and therefore includes repeated use of an utterance in MMLU-SIB.}
\label{tab:app-sib-construction}
\setlength{\tabcolsep}{12pt}
\renewcommand{\arraystretch}{1.10}
\begin{tabular}{lrr}
\toprule
\rowcolor{TableHeader}
\textbf{Statistic} & \textbf{ARC-SIB} & \textbf{MMLU-SIB} \\
\midrule
Source speakers & 40 & 40 \\
\rowcolor{TableAlt}
Source chapters & 97 & 97 \\
Source duration & 5.39 h & 5.39 h \\
\rowcolor{TableAlt}
Text--speech pairs & 1,172 & 14,042 \\
Unique speech utterances & 1,172 & 2,694 \\
\rowcolor{TableAlt}
Mean utterance reuse & 1.000 & 5.212 \\
Median / p90 / p95 reuse & 1 / 1 / 1 & 5 / 8 / 9 \\
\rowcolor{TableAlt}
Maximum utterance reuse & 1 & 14 \\
Processed duration per pair & 5 s & 5 s \\
\rowcolor{TableAlt}
Total processed paired duration & 1.63 h & 19.50 h \\
\bottomrule
\end{tabular}
\end{table}

The two tasks draw from a common speech pool. A total of 1,170 utterances are shared across ARC-SIB and MMLU-SIB, and their union contains 2,696 unique utterances. Thus, 99.83\% of ARC-SIB utterances also occur in MMLU-SIB. The two conditions evaluate different text tasks under related speech pairings.

\subsection{Construction Protocol}
\label{app:sib-construction}

The construction scripts use fixed-duration five-second clips. Long utterances are randomly cropped and short utterances are zero-padded. ARC-SIB samples without replacement, whereas MMLU-SIB samples with replacement. The recorded construction defaults to seed zero and omits RMS normalization. The dataset files and construction scripts support recovery of speaker, chapter, and utterance identifiers from standard LibriSpeech paths. Unrecorded command-line overrides and original crop-offset distributions are excluded from the reported claims.

\subsection{Transcript-Level Validity Audit}
\label{app:sib-transcript-audit}

To test whether random speech pairings accidentally contain question-related or answer-bearing content, we recover transcripts for all 14,042 MMLU-SIB pairs and apply the existing IDF overlap screen. The screen flags 28 pairs, corresponding to 0.20\% of the benchmark. Manual review classifies 22 flagged pairs as lexical false positives and six as weakly topical; none is judged potentially relevant to the answer and none contains answer-bearing speech. Table~\ref{tab:app-sib-validity} summarizes this audit.

\begin{table}[t]
\centering
\small
\caption{MMLU-SIB transcript-level validity audit. The audit checks the constructed speech--question pairings and is not part of the ICAP-Gate routing rule.}
\label{tab:app-sib-validity}
\begin{tabular}{lr}
\toprule
\rowcolor{TableHeader}
\textbf{Audit item} & \textbf{Count or rate} \\
\midrule
MMLU-SIB pairs & 14,042 \\
Pairs with recovered transcript & 14,042 \\
Missing transcripts & 0 \\
IDF candidates above threshold & 28 (0.20\%) \\
Lexical false positives among candidates & 22 \\
Weakly topical among candidates & 6 \\
\rowcolor{TablePositive}
Potentially relevant among candidates & \textbf{0} \\
\rowcolor{TablePositive}
Answer-bearing speech among candidates & \textbf{0} \\
\bottomrule
\end{tabular}
\end{table}

The audit supports \texttt{speech\_random} as an irrelevant-speech condition for MMLU-SIB. The condition targets random single-speaker speech; conversational, multi-speaker, semantically conflicting, language-controlled, and signal-to-noise-controlled interference remain outside this evaluation.

\subsection{Speech-Pairing Summary}
\label{app:sib-boundaries}

The pairing is traceable to a defined LibriSpeech pool, and the full MMLU-SIB audit finds no answer-bearing speech among screened candidates. MMLU-SIB reuses utterances from the same pool as ARC-SIB, allowing the two conditions to evaluate different text tasks under related speech pairings. The recorded source indices support the quantitative analyses reported for both conditions.

\end{document}

%% file: math_commands.tex
\usepackage{amsmath,amsfonts,bm}

\def\eqref#1{equation~\ref{#1}}

\def\1{\bm{1}}

\DeclareMathAlphabet{\mathsfit}{\encodingdefault}{\sfdefault}{m}{sl}
\SetMathAlphabet{\mathsfit}{bold}{\encodingdefault}{\sfdefault}{bx}{n}



%% file: iclr2027_conference.bib
@inproceedings{deshmukh2023pengi,
  title     = {{Pengi}: An Audio Language Model for Audio Tasks},
  author    = {Deshmukh, Soham and Elizalde, Benjamin and Singh, Rita and Wang, Huaming},
  booktitle = {Advances in Neural Information Processing Systems},
  volume    = {36},
  pages     = {18090--18108},
  year      = {2023}
}

@inproceedings{gong2024listen,
  title     = {Listen, Think, and Understand},
  author    = {Gong, Yuan and Luo, Hongyin and Liu, Alexander H. and Karlinsky, Leonid and Glass, James R.},
  booktitle = {International Conference on Learning Representations},
  year      = {2024}
}

@inproceedings{zhang2023speechgpt,
  title={Speechgpt: Empowering large language models with intrinsic cross-modal conversational abilities},
  author={Zhang, Dong and Li, Shimin and Zhang, Xin and Zhan, Jun and Wang, Pengyu and Zhou, Yaqian and Qiu, Xipeng},
  booktitle={Findings of the Association for Computational Linguistics: EMNLP 2023},
  pages={15757--15773},
  year={2023}
}

@article{rubenstein2023audiopalm,
  title={{AudioPaLM}: A large language model that can speak and listen},
  author={Rubenstein, Paul K and Asawaroengchai, Chulayuth and Nguyen, Duc Dung and Bapna, Ankur and Borsos, Zal{\'a}n and Quitry, F{\'e}lix de Chaumont and Chen, Peter and Badawy, Dalia El and Han, Wei and Kharitonov, Eugene and others},
  journal={arXiv preprint arXiv:2306.12925},
  year={2023}
}

@inproceedings{tang2024salmonn,
  title     = {{SALMONN}: Towards Generic Hearing Abilities for Large Language Models},
  author    = {Tang, Changli and Yu, Wenyi and Sun, Guangzhi and Chen, Xianzhao and Tan, Tian and Li, Wei and Lu, Lu and Ma, Zejun and Zhang, Chao},
  booktitle = {International Conference on Learning Representations},
  year      = {2024}
}

@article{chu2023qwen,
  title={{Qwen-Audio}: Advancing universal audio understanding via unified large-scale audio-language models},
  author={Chu, Yunfei and Xu, Jin and Zhou, Xiaohuan and Yang, Qian and Zhang, Shiliang and Yan, Zhijie and Zhou, Chang and Zhou, Jingren},
  journal={arXiv preprint arXiv:2311.07919},
  year={2023}
}

@article{chu2024qwen2,
  title={{Qwen2-Audio} technical report},
  author={Chu, Yunfei and Xu, Jin and Yang, Qian and Wei, Haojie and Wei, Xipin and Guo, Zhifang and Leng, Yichong and Lv, Yuanjun and He, Jinzheng and Lin, Junyang and others},
  journal={arXiv preprint arXiv:2407.10759},
  year={2024}
}

@inproceedings{ghosh2025audio,
  title     = {{Audio Flamingo 2}: An Audio-Language Model with Long-Audio Understanding and Expert Reasoning Abilities},
  author    = {Ghosh, Sreyan and Kong, Zhifeng and Kumar, Sonal and Sakshi, S and Kim, Jaehyeon and Ping, Wei and Valle, Rafael and Manocha, Dinesh and Catanzaro, Bryan},
  booktitle = {Proceedings of the 42nd International Conference on Machine Learning},
  pages     = {19358--19405},
  year      = {2025},
  volume    = {267},
  series    = {Proceedings of Machine Learning Research},
  publisher = {PMLR}
}

@article{defossez2024moshi,
  title={Moshi: a speech-text foundation model for real-time dialogue},
  author={D{\'e}fossez, Alexandre and Mazar{\'e}, Laurent and Orsini, Manu and Royer, Am{\'e}lie and P{\'e}rez, Patrick and J{\'e}gou, Herv{\'e} and Grave, Edouard and Zeghidour, Neil},
  journal={arXiv preprint arXiv:2410.00037},
  year={2024}
}

@article{huang2025step,
  title={Step-audio: Unified understanding and generation in intelligent speech interaction},
  author={Huang, Ailin and Wu, Boyong and Wang, Bruce and Yan, Chao and Hu, Chen and Feng, Chengli and Tian, Fei and Shen, Feiyu and Li, Jingbei and Chen, Mingrui and others},
  journal={arXiv preprint arXiv:2502.11946},
  year={2025}
}

@article{li2025baichuan,
  title={Baichuan-omni-1.5 technical report},
  author={Li, Yadong and Liu, Jun and Zhang, Tao and Chen, Song and Li, Tianpeng and Li, Zehuan and Liu, Lijun and Ming, Lingfeng and Dong, Guosheng and Pan, Da and others},
  journal={arXiv preprint arXiv:2501.15368},
  year={2025}
}

@article{xu2025qwen25omnitechnicalreport,
  title   = {Qwen2.5-Omni Technical Report},
  author  = {Xu, Jin and Guo, Zhifang and He, Jinzheng and Hu, Hangrui and He, Ting and Bai, Shuai and Chen, Keqin and Wang, Jialin and Fan, Yang and Dang, Kai and Zhang, Bin and Wang, Xiong and Chu, Yunfei and Lin, Junyang},
  journal = {arXiv preprint arXiv:2503.20215},
  year    = {2025}
}

@article{abouelenin2025phi,
  title={Phi-4-mini technical report: Compact yet powerful multimodal language models via mixture-of-loras},
  author={Abouelenin, Abdelrahman and Ashfaq, Atabak and Atkinson, Adam and Awadalla, Hany and Bach, Nguyen and Bao, Jianmin and Benhaim, Alon and Cai, Martin and Chaudhary, Vishrav and Chen, Congcong and others},
  journal={arXiv preprint arXiv:2503.01743},
  year={2025}
}

@article{liu2025voxtral,
  title={Voxtral},
  author={Liu, Alexander H and Ehrenberg, Andy and Lo, Andy and Denoix, Cl{\'e}ment and Barreau, Corentin and Lample, Guillaume and Delignon, Jean-Malo and Chandu, Khyathi Raghavi and von Platen, Patrick and Muddireddy, Pavankumar Reddy and others},
  journal={arXiv preprint arXiv:2507.13264},
  year={2025}
}

@inproceedings{shi2023large,
  title={Large language models can be easily distracted by irrelevant context},
  author={Shi, Freda and Chen, Xinyun and Misra, Kanishka and Scales, Nathan and Dohan, David and Chi, Ed H and Sch{\"a}rli, Nathanael and Zhou, Denny},
  booktitle={International conference on machine learning},
  pages={31210--31227},
  year={2023},
  organization={PMLR}
}

@inproceedings{yoran2024making,
  title={Making retrieval-augmented language models robust to irrelevant context},
  author={Yoran, Ori and Wolfson, Tomer and Ram, Ori and Berant, Jonathan},
  booktitle={International Conference on Learning Representations},
  volume={2024},
  pages={29862--29883},
  year={2024}
}

@inproceedings{yang2025llm,
  title={How is llm reasoning distracted by irrelevant context? an analysis using a controlled benchmark},
  author={Yang, Minglai and Huang, Ethan and Zhang, Liang and Surdeanu, Mihai and Wang, William Yang and Pan, Liangming},
  booktitle={Proceedings of the 2025 Conference on Empirical Methods in Natural Language Processing},
  pages={13340--13358},
  year={2025}
}

@inproceedings{sharma2024losing,
  title={Losing visual needles in image haystacks: Vision language models are easily distracted in short and long contexts},
  author={Sharma, Aditya and Saxon, Michael and Wang, William Yang},
  booktitle={Findings of the Association for Computational Linguistics: EMNLP 2024},
  pages={5429--5451},
  year={2024}
}

@article{yang2026defying,
  title={Defying Distractions in Multimodal Tasks: A Novel Benchmark for Large Vision-Language Models},
  author={Yang, Jinhui and Jiang, Ming and Zhao, Qi},
  journal={IEEE Transactions on Pattern Analysis and Machine Intelligence},
  year={2026},
  publisher={IEEE}
}

@article{cai2025diagnosing,
  title={Diagnosing and mitigating modality interference in multimodal large language models},
  author={Cai, Rui and Li, Bangzheng and Wen, Xiaofei and Chen, Muhao and Zhao, Zhe},
  journal={arXiv preprint arXiv:2505.19616},
  year={2025}
}

@article{liu2024insight,
  title={Insight over sight: Exploring the vision-knowledge conflicts in multimodal LLMs},
  author={Liu, Xiaoyuan and Wang, Wenxuan and Yuan, Youliang and Huang, Jen-tse and Liu, Qiuzhi and He, Pinjia and Tu, Zhaopeng},
  journal={arXiv preprint arXiv:2410.08145},
  year={2024}
}

@inproceedings{tian2026crosscheck,
  title={Crosscheck-bench: Diagnosing compositional failures in multimodal conflict resolution},
  author={Tian, Baoliang and Si, Yuxuan and Wang, Jilong and Li, LingYao and Bao, Zhongyuan and Zhou, Zineng and Wang, Tao and Li, Sixu and Xu, Ziyao and Wang, Mingze and others},
  booktitle={Proceedings of the AAAI Conference on Artificial Intelligence},
  volume={40},
  pages={25887--25895},
  year={2026}
}

@inproceedings{li2026silence,
  title={When silence matters: The impact of irrelevant audio on text reasoning in large audio-language models},
  author={Li, Chen-An and Lin, Tzu-Han and Lee, Hung-yi},
  booktitle={ICASSP 2026-2026 IEEE International Conference on Acoustics, Speech and Signal Processing (ICASSP)},
  pages={17757--17761},
  year={2026},
  organization={IEEE}
}

@article{vig2020investigating,
  title={Investigating gender bias in language models using causal mediation analysis},
  author={Vig, Jesse and Gehrmann, Sebastian and Belinkov, Yonatan and Qian, Sharon and Nevo, Daniel and Singer, Yaron and Shieber, Stuart},
  journal={Advances in neural information processing systems},
  volume={33},
  pages={12388--12401},
  year={2020}
}

@article{meng2022locating,
  title={Locating and editing factual associations in gpt},
  author={Meng, Kevin and Bau, David and Andonian, Alex and Belinkov, Yonatan},
  journal={Advances in neural information processing systems},
  volume={35},
  pages={17359--17372},
  year={2022}
}

@article{conmy2023towards,
  title={Towards automated circuit discovery for mechanistic interpretability},
  author={Conmy, Arthur and Mavor-Parker, Augustine and Lynch, Aengus and Heimersheim, Stefan and Garriga-Alonso, Adri{\`a}},
  journal={Advances in Neural Information Processing Systems},
  volume={36},
  pages={16318--16352},
  year={2023}
}

@inproceedings{zhang2024towards,
  title     = {Towards Best Practices of Activation Patching in Language Models: Metrics and Methods},
  author    = {Zhang, Fred and Nanda, Neel},
  booktitle = {International Conference on Learning Representations},
  year      = {2024}
}

@inproceedings{ghandeharioun2024patchscopes,
  title     = {{Patchscopes}: A Unifying Framework for Inspecting Hidden Representations of Language Models},
  author    = {Ghandeharioun, Asma and Caciularu, Avi and Pearce, Adam and Dixon, Lucas and Geva, Mor},
  booktitle = {Proceedings of the 41st International Conference on Machine Learning},
  pages     = {15466--15490},
  year      = {2024},
  volume    = {235},
  series    = {Proceedings of Machine Learning Research},
  publisher = {PMLR}
}

@inproceedings{palit2023towards,
  title     = {Towards Vision-Language Mechanistic Interpretability: A Causal Tracing Tool for {BLIP}},
  author    = {Palit, Vedant and Pandey, Rohan and Arora, Aryaman and Liang, Paul Pu},
  booktitle = {Proceedings of the IEEE/CVF International Conference on Computer Vision Workshops},
  pages     = {2856--2861},
  year      = {2023}
}

@inproceedings{li2026causal,
  title={Causal tracing of object representations in large vision language models: Mechanistic interpretability and hallucination mitigation},
  author={Li, Qiming and Ye, Zekai and Feng, Xiaocheng and Zhong, Weihong and Ma, Weitao and Feng, Xiachong},
  booktitle={Proceedings of the AAAI Conference on Artificial Intelligence},
  volume={40},
  pages={31645--31653},
  year={2026}
}

@inproceedings{kim2026map,
  title     = {Map the Flow: Revealing Hidden Pathways of Information in {VideoLLMs}},
  author    = {Kim, Minji and Kim, Taekyung and Han, Bohyung},
  booktitle = {International Conference on Learning Representations},
  year      = {2026}
}

@inproceedings{arevalo2017gated,
  title     = {Gated Multimodal Units for Information Fusion},
  author    = {Arevalo, John and Solorio, Thamar and Montes-y-G{\'o}mez, Manuel and Gonz{\'a}lez, Fabio A.},
  booktitle = {International Conference on Learning Representations},
  year      = {2017}
}

@inproceedings{perez2018film,
  title={Film: Visual reasoning with a general conditioning layer},
  author={Perez, Ethan and Strub, Florian and De Vries, Harm and Dumoulin, Vincent and Courville, Aaron},
  booktitle={Proceedings of the AAAI conference on artificial intelligence},
  volume={32},
  year={2018}
}

@article{alayrac2022flamingo,
  title={Flamingo: a visual language model for few-shot learning},
  author={Alayrac, Jean-Baptiste and Donahue, Jeff and Luc, Pauline and Miech, Antoine and Barr, Iain and Hasson, Yana and Lenc, Karel and Mensch, Arthur and Millican, Katherine and Reynolds, Malcolm and others},
  journal={Advances in neural information processing systems},
  volume={35},
  pages={23716--23736},
  year={2022}
}

@inproceedings{shazeer2017outrageously,
  title     = {Outrageously Large Neural Networks: The Sparsely-Gated Mixture-of-Experts Layer},
  author    = {Shazeer, Noam and Mirhoseini, Azalia and Maziarz, Krzysztof and Davis, Andy and Le, Quoc and Hinton, Geoffrey and Dean, Jeff},
  booktitle = {International Conference on Learning Representations},
  year      = {2017}
}

@article{fedus2022switch,
  title={Switch transformers: Scaling to trillion parameter models with simple and efficient sparsity},
  author={Fedus, William and Zoph, Barret and Shazeer, Noam},
  journal={Journal of Machine Learning Research},
  volume={23},
  number={120},
  pages={1--39},
  year={2022}
}

@inproceedings{wu2024omni,
  title={Omni-smola: Boosting generalist multimodal models with soft mixture of low-rank experts},
  author={Wu, Jialin and Hu, Xia and Wang, Yaqing and Pang, Bo and Soricut, Radu},
  booktitle={2024 IEEE/CVF Conference on Computer Vision and Pattern Recognition (CVPR)},
  pages={14205--14215},
  year={2024},
  organization={IEEE}
}

@article{lin2026moe,
  title     = {{MoE-LLaVA}: Mixture of Experts for Large Vision-Language Models},
  author    = {Lin, Bin and Tang, Zhenyu and Ye, Yang and Huang, Jinfa and Zhang, Junwu and Pang, Yatian and Jin, Peng and Ning, Munan and Luo, Jiebo and Yuan, Li},
  journal   = {IEEE Transactions on Multimedia},
  volume    = {28},
  pages     = {4408--4419},
  year      = {2026},
  publisher = {IEEE}
}

@article{fonseca2022fsd50k,
  title={{FSD50K}: An Open Dataset of Human-Labeled Sound Events},
  author={Fonseca, Eduardo and Favory, Xavier and Pons, Jordi and Font, Frederic and Serra, Xavier},
  journal={IEEE/ACM Transactions on Audio, Speech, and Language Processing},
  volume={30},
  pages={829--852},
  year={2022},
  publisher={IEEE}
}

@inproceedings{panayotov2015librispeech,
  title={{LibriSpeech}: An {ASR} Corpus Based on Public Domain Audio Books},
  author={Panayotov, Vassil and Chen, Guoguo and Povey, Daniel and Khudanpur, Sanjeev},
  booktitle={2015 IEEE International Conference on Acoustics, Speech and Signal Processing (ICASSP)},
  pages={5206--5210},
  year={2015},
  organization={IEEE}
}
